# Integrated Modeling of Second Phase Precipitation in Cold-Worked 316 Stainless Steels under Irradiation

Mahmood Mamivand[1], Ying Yang[2], Jeremy Busby[2], Dane Morgan[1*]

[1] Department of Materials Science and Engineering, University of Wisconsin-Madison
[2] Materials Science & Technology Division, Oak Ridge National Laboratory
[*]Corresponding Author: ddmorgan@wisc.edu, Tel: (608) 265-5879

**Abstract**

The current work combines the Cluster Dynamics (CD) technique and CALPHAD-based precipitation modeling to address the second phase precipitation in cold-worked (CW) 316 stainless steels (SS) under irradiation at 300-400 °C. CD provides the radiation enhanced diffusion and dislocation evolution as inputs for the precipitation model. The CALPHAD-based precipitation model treats the nucleation, growth and coarsening of precipitation processes based on classical nucleation theory and evolution equations, and simulates the composition, size and size distribution of precipitate phases. We benchmark the model against available experimental data at fast reactor conditions ($9.4 \times 10^{-7}$ dpa/s and 390 °C) and then use the model to predict the phase instability of CW 316 SS under light water reactor (LWR) extended life conditions ($7 \times 10^{-8}$ dpa/s and 275 °C). The model accurately predicts the $\gamma'$ (Ni$_3$Si) precipitation evolution under fast reactor conditions and that the formation of this phase is dominated by radiation enhanced segregation. The model also predicts a carbide volume fraction that agrees well with available experimental data from a PWR reactor but is much higher than the volume fraction observed in fast reactors. We propose that radiation enhanced dissolution and/or carbon depletion at sinks that occurs at high flux could be the main sources of this inconsistency. The integrated model predicts ~1.2% volume fraction for carbide and ~3.0% volume fraction for $\gamma'$ for typical CW 316 SS (with 0.054 wt.% carbon) under LWR extended life conditions. This work provides valuable insights into the magnitudes and mechanisms of precipitation in irradiated CW 316 SS for nuclear applications.

***Keywords:*** *Austenitic stainless steels, Radiation induced precipitation, Radiation enhanced*

1. Introduction

Austenitic stainless steels (SS) are major structural materials in a reactor core because of their high strength, corrosion resistance, and formability. In a reactor core, structural materials experience a relatively harsh environment (> 250 °C and neutron irradiation), which leads to materials degradation. One form of austenitic SS degradation is precipitation of second phases in the matrix. Precipitation in austenitic SS under irradiation could lead to steel hardening and embrittlement and finally limiting their operation lifetime [1-3].

Precipitation in austenitic SS under irradiation has been subject of several experimental studies especially for temperatures higher than 400 °C [4-8]. It was thought for some years that no significant precipitation may happen in austenitic SS at temperatures lower than 400 °C [5]. However, later experimental observations showed that phases like carbides ($M_{23}C_6$ and $M_6C$), $\gamma'$ ($Ni_3Si$), and G-phase ($M_6Ni_{16}Si_7$) could form in austenitic SS under irradiation at temperature lower than 400 °C and specifically in the temperature range experienced by todays Light Water Reactor (LWR) core-internal components (275-340 °C) [9-13].

Most of the available experimental data come from irradiated specimens at fast reactors and the precipitation in austenitic SS under LWR conditions is less well studied. The possibility of life limiting effects of precipitates becomes a particular concern under LWR extended life conditions, as there is limited experimental data to address the materials behavior under such low-flux, high-fluence conditions. In the absence of experimental data, modeling techniques can help us to gain insight into materials behavior under LWR extended life conditions.

There are two recent works with an emphasis on thermodynamic and kinetic modeling of austenitic SS precipitates under nuclear power plants conditions. Yang and Busby [14] used the CALPHAD (Calculation of Phase Diagram) approach and developed a thermodynamic database, OCTANT (ORNL Computational Thermodynamics for Applied Nuclear Technology), for austenitic stainless steels with a focus on reliable thermodynamic modeling of precipitate phases in AISI 316. Then they coupled the thermodynamic database with precipitation kinetics simulation (using MatCalc package [15]) to study the thermal aging of 316 SS. They found that by increasing dislocation density, the precipitation kinetics of phases such as $M_{23}C_6$ and $Ni_3Si$ can be greatly enhanced. Shim et al. [16] used the same methodology (CALPHAD database and MatCalc) to study the thermal aging of 316 SS at 400 °C. They also studied the aging behavior of alloys with the addition of radiation induced segregation (RIS) composition to gain insight into radiation induced precipitates (RIP), and demonstrated that RIS can play a critical role in 316 SS precipitates. In both works authors commented on the lack of correct radiation enhanced diffusion (RED) and proper dislocation evolution in their modeling as the key missing information which held them back from more accurate modeling of 316 SS under irradiation. A valid estimation of RED is critical in obtaining correct time scales of precipitation as the diffusion controls many aspects of precipitate growth. Dislocations evolution is also essential to a precipitation model for multiple reasons. First, to develop an accurate RED model a realistic dislocation density is necessary, as it provides a sink density in the defect evolution equations. Furthermore, dislocations will influence the RIS through their impact on overall sink density and their ability to generate RIS near the dislocation core. Finally, dislocations can serve as fast diffusion channels and nucleation sites for precipitates. We seek to integrate RED and dislocation models into precipitate evolution models to gain insight into the mechanisms, time scales, and extent of precipitations in CW 316 SS.

 To address the aspects lacking in previous works we developed an integrated model which combines the Cluster Dynamics (CD), precipitation modeling (MatCalc), thermodynamic database (OCTANT), and mobility database (mc_fe_v2.006). We use multiple programs as no one set of codes provides all the necessary modeling tools. The CD code was written ourselves and tracks the evolution of defects under irradiation and gives us the evolution of single vacancy (and consequently RED) and total dislocation concentrations. The CD predicted RED and dislocation density go into MatCalc as input data. In MatCalc the evolution of the thermodynamic system is based on the framework of the Kampmann–Wagner model [17] which breaks the total time history into small isothermal segments. The nucleation kinetics is based on classical nucleation theory extended for multicomponent systems [18] and the growth is evaluated based on the evolution equations for the radius and composition of the precipitate derived by Svoboda et al. [18] which is based on a mean-field approach utilizing the thermodynamic extremal principle [19, 20]. More detail reading about MatCalc can be found in [21].

 For radiation induced phases (RIPs), which occur primarily around dislocation loops, we use a simple model that assumes that RIPs form inside a cylindrical region around the dislocations loops, where this region is assumed to have the appropriate RIS composition. Using the RIS composition to study the RIP is similar to the approach used by Shim et al. [16]. However, in this work, we use more quantitative a cylindrical model to address the RIP (details are given in section 3.2.4).

## 2. Methods

### 2.1 Cluster dynamics: governing equations

Cluster Dynamics (CD) is a computational technique for predicting microstructural evolution and it is frequently applied to precipitation problems or defect cluster evolution in materials under irradiation. In CD, the system is described as a gas of non-interacting clusters. The clusters are defined by a single parameter, their size (or equivalently, the number of atoms they contain).

In CD modeling of defect clusters the principle of the model is to describe a population of defects by their size distribution. The evolution of these populations is obtained through 'chemical kinetics' in a homogeneous medium, where the probability of a cluster of size $n$ to become a cluster of size $n+1$ or $n-1$ depends on its rate of absorption or emission of a vacancy or an interstitial. These kinetics depend on the available population of mobile defects.

The main parts of the CD model developed here are:
- Rate of defect production from irradiation cascade,
- Recombination rate of point defects,
- Absorption and emission rates of point defects by the defect clusters (loops and voids),
- Annihilation kinetics on fixed sinks like grain boundaries,
- Annihilation kinetics on dislocations,
- Frank loops un-faulting,
- Network dislocation evolution.

The model contains a series of coupled ordinary differential equations that capture the evolution of point defects and larger clusters. The solution of these equations is obtained by direct integration of equations using the CVODE solver [22]. Our approach to building this model will be to use existing models and then alter them as needed to yield agreement with known data on loop evolution.

The modeling approach is taken from references [23-29]. The generation of defects from the cascade is taken from [30] which considers the formation of clusters of size higher than 4 unlikely. The defect generation terms are

$$G_{i(v)}(1) = \eta G_{dpa}(1 - f_{i(v)2} - f_{i(v)3} - f_{i(v)4}),$$
$$G_{i(v)}(2) = \frac{\eta G_{dpa} f_{i(v)2}}{2}, \tag{1}$$

$$G_{i(v)}(3) = \frac{\eta G_{dpa} f_{i(v)3}}{3},$$

$$G_{i(v)}(4) = \frac{\eta G_{dpa} f_{i(v)4}}{4},$$

$$G_{i(v)}(n > 4) = 0.$$

$G_{dpa}$ in these equations is the damage rate in the reactor, $\eta$ is cascade efficiency and $f_{i(v)n}$ is the fraction of clusters on size $n$ and type $i(v)$ surviving the reorganization events following the cascade.

Assuming that only monomer defects are mobile the governing equations for defect evolution would be as follows [23-27]

$$\frac{dC_{i(v)}(1)}{dt} = G_{i(v)}(1) - R_{iv}C_i(1)C_v(1) - \rho D_{i(v)} Z_{i(v)}[C_{i(v)}(1) - C^e_{i(v)}] - 6D_{i(v)} \frac{\sqrt{S^{i(v)}_m}}{d_g}[C_{i(v)}(1) - C^e_{i(v)}]$$
$$-[\sum_{n=2} \beta_{i(v),i(v)}(n)C_{i(v)}(n) + \sum_{n=2} \beta_{v(i),i(v)}(n)C_{v(i)}(n) + 4\beta_{i(v),i(v)}(1)C_{i(v)}(1)]C_{i(v)}(1) +$$
$$\sum_{n=3} \alpha_{i(v),i(v)}(n)C_{i(v)}(n) + 4\alpha_{i(v),i(v)}(2)C_{i(v)}(2) + \beta_{i(v),v(i)}(2)C_{i(v)}(2)C_{v(i)}(1)$$
(2)

$$\frac{dC_{i(v)}(2)}{dt} = G_{i(v)}(2) + 2\beta_{i(v),i(v)}(1)C_{i(v)}(1)C_{i(v)}(1) - 2\alpha_{i(v),i(v)}(2)C_{i(v)}(2) -$$
$$\beta_{i(v),i(v)}(2)C_{i(v)}(1)C_{i(v)}(2) + \alpha_{i(v),i(v)}(3)C_{i(v)}(3) - \beta_{i(v),v(i)}(2)C_{v(i)}(1)C_{i(v)}(2)$$
$$+ \beta_{i(v),v(i)}(3)C_{v(i)}(1)C_{i(v)}(3)$$
(3)

$$\frac{dC_{i(v)}(n)}{dt} = G_{i(v)}(n) + [\beta_{i(v),v(i)}(n+1)C_{v(i)}(1) + \alpha_{i(v),i(v)}(n+1)]C_{i(v)}(n+1) -$$
$$[\beta_{i(v),v(i)}(n)C_{v(i)}(1) + \beta_{i(v),i(v)}(n)C_{i(v)}(1) + \alpha_{i(v),i(v)}(n)]C_{i(v)}(n) +$$
$$[\beta_{i(v),i(v)}(n-1)C_{i(v)}(1)]C_{i(v)}(n-1) \qquad n > 2$$
(4)

here $i$ and $v$ refer to interstitials and vacancies (either as point defects or cluster types (i.e., loops and voids, respectively)), the $C_\theta(n)$ is the concentration per lattice site of clusters of type $\theta$ ($\theta$ can be type interstitial ($i$) or vacancy ($v$)) containing $n$ atoms, $G_\theta(n)$ is the production rate of cluster of size $n$, $R_{iv}$ is the characteristic annihilation rate of vacancy and interstitial, $\rho$ is the background dislocation density, $D_{i(v)}$ is the interstitial (vacancy) diffusion coefficient, $Z_{i(v)}$ is the interstitial (vacancy) capture efficiency by dislocation net, $C^e_{i(v)}$ is the thermal equilibrium concentration of interstitial (vacancy), $d_g$ is the grain boundary size, $S_m^{i(v)}$ is the characteristic grain boundary sink strength, $\beta_{\theta,\theta'}(n)$ and $\alpha_{\theta,\theta'}(n)$ are the rate of absorption and emission of a defect of type $\theta'$ by a cluster of type $\theta$ and size $n$ respectively. $D_{i(v)}$, $R_{iv}$, $S_m^{i(v)}$, $\beta_{\theta,\theta'}(n)$, and $\alpha_{\theta,\theta'}(n)$ are defined as following,

$$D_{i(v)} = D_{0i(v)} \exp(-\frac{E_{mi(v)}}{kT}), \tag{5}$$

$$R_{iv} = 4\pi(D_v + D_i)r_{iv}/V_{at}, \tag{6}$$

$$S_m^{i(v)} = \rho Z_{i(v)} + \frac{1}{D_{i(v)}} \sum_{n=2}(\beta_{i(v),i(v)}(n)C_{i(v)}(n) + \beta_{v(i),i(v)}(n)C_{v(i)}(n)), \tag{7}$$

$$\beta_{i,v}(n) = \frac{2\pi r_i(n)D_v Z_v(n)}{V_{at}}, \tag{8}$$

$$\beta_{i,i}(n) = \frac{2\pi r_i(n)D_i Z_i(n)}{V_{at}}, \tag{9}$$

$$\beta_{v,i}(n) = \frac{2\pi r_v(n)D_i Z_i(n)}{V_{at}}, \tag{10}$$

$$\beta_{v,v}(n) = \frac{2\pi r_v(n)D_v Z_v(n)}{V_{at}}, \tag{11}$$

$$\alpha_{i,i}(n) = \beta_{i,i}(n-1)\exp(-\frac{E_{bi}(n)}{kT}), \tag{12}$$

$$\alpha_{v,v}(n) = \beta_{v,v}(n-1)\exp(-\frac{E_{bv}(n)}{kT}). \tag{13}$$

In above equations the $D_{0i(v)}$ is interstitial (vacancy) pre-exponential, $E_{mi(v)}$ is interstitial (vacancy) migration energy, $k$ is Boltzmann constant, $T$ is the temperature, $r_{iv}$ is interstitial-vacancy capture distance, $V_{at}$ is the average atomic volume of the steel, $r_\theta(n)$ is the size of the cluster of type $\theta$ containing $n$ point defects,. Critical parameters in the above equations (8)-(11) are the bias factors $Z_\theta(n)$ of cluster of size $n$. This bias for interstitial clusters can be defined as [24],

$$Z_{i(v)}(n) = Z_{i(v)} + (\sqrt{\frac{b}{8\pi a}}Z_{1i(v)} - Z_{i(v)})(1/n^{0.35}), \tag{14}$$

where $Z_{i(v)}$ is the bias factor for an infinite straight dislocation for the interstitial (vacancy) point defects, $a$ is the lattice parameter, $b$ is the Burgers vector and $Z_{1i(v)}$ is a parameter used to describe the evolution of the bias $Z_{i(v)}(n)$ with the size of the clusters.

The equations (12) and (13), which represent the parameters controlling emission in the CD model, are highly dependent on binding energy. Based on molecular dynamics simulation the binding energies for interstitials in iron can be described by the following expression [31, 32],

$$E_{bi(v)}(n) = E_{fi(v)} + \frac{E_{b2i(v)} - E_{fi(v)}}{2^{2/3} - 1}(n^{2/3} - (n-1)^{2/3}), \tag{15}$$

where $E_{fi(v)}$ is the formation energy of interstitial (vacancy) point defects, $n$ is the number of atom in cluster, and $E_{b2i}$ is the binging energy for a cluster of size two. We will use this parameterization for the present austenitic systems as well.

Another important phenomenon that needs to be considered in modeling loop evolution is the evolution of network dislocations. Network dislocations are both annihilated and produced under high temperature irradiation. The annihilation comes from the dislocation climb due to excess defects presence under irradiation, and the production comes from a high temperature climb source, also known as Bardeen-Herring source, and Frank loops un-faulting. A Bardeen-Herring dislocation source is similar to a Frank-Reed source except the former is climb driven while the latter is glide driven [33]. The other source, loop un-faulting, is generally thought to occur once a given loop grows sufficiently large that it intersects another microstructure feature. This intersection can generate enough fluctuation to nucleate an un-faulting Shockley partial which glides in the Frank loop plane and transforms the Frank loop into a perfect loop [34]. The aforementioned production and annihilation processes can be introduced as source and sink terms, respectively, into a model of the evolution of network dislocation density, $\rho$. This model, based on references [28, 29], gives the evolution of network dislocation density as follows,

$$\frac{d\rho}{dt} = 2\pi v_{cl} S_{BH} - \frac{\rho}{\tau_{cl}} + 2\pi \sum_{n \geq 2} \frac{r_i(n) C_i(n)}{\tau_i(n) . V_{at}}, \tag{16}$$

where the first term on the right hand side is Bardeen-Herring source term, the second term is climb annihilation term, and the third term is loop un-faulting source term.

In equation (16) the $\rho$ is the density of network dislocation, $v_{cl}$ is the dislocation climb velocity, $S_{BH}$ is the Bardeen-Herring source density, $\tau_{cl}$ is the mean lifetime against annihilation due climb, $r_i(n)$ is the radius of a loop containing $n$ atoms, and $\tau_i(n)$ is the time necessary for the loop of size $n$ to incorporate into the network dislocation. Dislocation climb velocity due to excess vacancies and interstitials under irradiation is:

$$v_{cl} = \frac{2\pi}{\ln(r_o/r_c)} \frac{1}{b}[Z_i D_i C_i(1) - Z_v D_v (C_v(1) - C_v^n)], \qquad (17)$$

where $r_c$ is the dislocation core radius, $r_o$ is the outer radius of the cylindrical cell used in calculating the dislocation sink strength, and $C_v^n$ is the concentration of vacancies in equilibrium with the dislocation [29]. $C_v^n$ is given by

$$C_v^n = C_v^e \exp(\frac{\sigma V_{at}}{kT}), \qquad (18)$$

$$\sigma = AGb\sqrt{\rho_p}, \qquad (19)$$

where $\sigma$ is internal (back) stress as the result of a population of immobilized dislocations, $A$ is a geometric parameter with a value of 0.4 based on Ref. [29], $G$ is the shear modulus, and $\rho_p$ is the density of pinned dislocation and is assumed to be a fraction (here 0.1) of the total network dislocation density [29]. The other terms in equation (16) can be calculate as,

$$S_{BH} = (\frac{\rho_p}{3})^{1.5}, \qquad (20)$$

$$\tau_{cl} = \frac{d_{cl}}{v_{cl}} = \frac{(\pi\rho)^{-1/2}}{v_{cl}}, \qquad (21)$$

$$\tau_i(n) = \frac{T_i^c(n)}{P_i^{unf}(n)}, \qquad (22)$$

$$T_i^c(n) = \frac{2\pi b r_i(n) a}{V_{at}[\beta_{i,i}(n)C_i(1) - \beta_{i,v}C_v(1) - \alpha_{i,i}(n)]}, \qquad (23)$$

$$P_i^{unf}(n) = C^{unf}[\sum_{m \geq 2} 2\pi r_i(m)C_i(m)/V_{at} + \rho]\pi_i(n). \qquad (24)$$

In equation (22), $T_i^c(n)$ is the time necessary for a loop of size $n$ to grow by $a$ (lattice parameter) and $P_i^{unf}(n)$ is the probability of loop of size $n$ to un-fault [28].

Interested readers are encouraged to refer to references [28, 29] for more details about network dislocation evolution model.

By solving the coupled master equations (2)-(4) and (16) one can capture the evolution of defects clusters during irradiation.

## 2.2 Cluster dynamics: parameters for 316 stainless steels

In CD modeling of defect clusters two sets of parameters are needed; 1) material parameters and 2) irradiation parameters. The goal is to find the best set of parameters that successfully reproduce the experimental data, in our case loop size and loop number density. Austenitic stainless steels have been the focus of several studies [26, 33, 35-37]. However, there is no established set of parameters for stainless steel CD models in the literature that cover all the complexity of defect cluster formation in these alloys. The material parameters used in this study are listed in Table 1. In this study we focus on CW 316 SS in order to have a concrete system for comparison. Similar approaches, likely with some tuning of parameters, can be adapted for other processing conditions for 316 SS and for 304 SS.

**Table 1.** Material parameters for 316 stainless steels.

| Parameter | Value | Reference |
|---|---|---|
| Lattice parameter, $a_0$ | 3.61 Å | |
| Interstitial migration energy, $E_{mi}$ | 0.43 eV | [26] |
| Vacancy migration energy, $E_{mv}$ | 1.35 eV | [26] |
| Interstitial pre-exponential, $D_{0i}$ | $1.0 \times 10^{-7}$ m²/s | [26] |
| Vacancy pre-exponential, $D_{0v}$ | $0.6 \times 10^{-4}$ m²/s | [26] |
| Interstitial formation energy, $E_{fi}$ | 4.1 eV | [26] |
| Vacancy formation energy, $E_{fv}$ | 1.61 eV | [38] |
| Vacancy formation entropy, $S_{fv}$ | $1.73\,k$ | [38] |
| Binding energy of interstitial dimer, $E_{b2i}$ | 0.5 eV | Fitting parameter |
| Binding energy of vacancy dimer, $E_{b2v}$ | 0.5 eV | [26] |
| Recombination radius, $r_{iv}$ | 0.7 nm | [26] |
| Dislocation density, $\rho_0$ | $1 \times 10^{14}$ m⁻² | [26] |
| Average grain size, $d$ | 40 µm | [26] |
| Burgers vector of the loop assumed to be prismatic, $b$ | $a/\sqrt{3}$ | [36] |
| Capture efficiency for interstitial by dislocation net, $Z_i$ | 1.25 | [36] |
| Capture efficiency for vacancy by dislocation net, $Z_v$ | 1.0 | [26] |
| $Z_{1i}$ | 42.0 | [26] |
| $Z_{1v}$ | 35.0 | [26] |
| $C^{unf}$ | 0.15 | [28] |

The other set of parameters are the irradiation parameters. These parameters characterize the irradiation conditions, which include the environmental parameters, e.g. temperature, damage rate and the in-cascade clustering behavior of the target material. The cascade properties might also be considered materials parameters, but because they may couple strongly to the irradiation conditions we consider them here as irradiation parameters. Irradiation parameters are reactor dependent and can vary from one reactor to the other depending on neutron flux and energy. Table 2 shows the irradiation parameters used in this work. Parameters for BOR-60 are obtained from reference [26]. For other reactors, the damage efficiencies are fitted to loop size and number density data reported in Ref. [10, 11, 37] in order to minimize the average of the absolute values of the errors between model predictions and experimental values. Note that these fits generally produce higher efficiencies at lower damage rates, consistent with expectations and trends used in Ref. [26]. Damage efficiencies were fitted by considering a grid of values every 0.05 in the range of 0.15 (corresponding to highest dose rate, BOR-60) and 0.5 (corresponding to 0.1 greater than the damage efficiency in reactor pressure vessels [39]).

In-cascade clustering parameters, $f_{ix(x=2,3,4)}$ and $f_{vx(x=2,3,4)}$, for LWR and Phénix reactor are selected with the following approaches. For LWR at 275 °C we initially used the in-cascade clustering parameters of BOR-60, but they produced more and smaller loops than experiments. Therefore, we reduced the interstitial in-cascade parameters with increment of 0.05 to get the minimum average absolute error between model predictions and experimental data in [37]. We did not alter the values of vacancy in-cascade clustering parameters ($f_{vx(x=2,3,4)}$) or $f_{i4}$ as they are all quite small and changing them had very little effect on loop size and number density. The best fit was obtained for $f_{i2} = 0.4$, $f_{i3} = 0.2$. For Phénix reactor we used the vacancy clustering parameters ($f_{vx}$) similar to EBR-II in Ref. [26] due to similarity in operating temperature. We fitted the interstitial clustering parameters ($f_{ix}$) to loop size and number density data in Ref. [11] by gridding the space of values from 0 to 1 in steps of 0.1 and taking those which gave the smallest absolute error vs. the experimental data. Finally, for LWR at 343 °C, we initially selected the parameters based on the values in [26], i.e., matching the values for LWR at 275 °C. However, the interstitial clustering parameters ($f_{ix}$) of LWR at 275 °C did not produce enough loop number density for LWR at 343 °C to match the data in Ref. [10]. Therefore, we used the Phénix interstitial clustering parameters, which were determined for similar temperatures and gave good agreement with experimental datum point in Ref. [10].

**Table 2.** Irradiation parameters used in CD model for different irradiation conditions.

| Parameter | BOR-60 [26] | Phénix [11] | LWR [10, 37] |
|---|---|---|---|
| Temperature, $T$ (°C) | 320 | 380, 381, 386 | 275, 343 |
| Dose rate, $G_{dpa}$ (dpa/s) | $9.4 \times 10^{-7}$ | $3.5 \times 10^{-7}$, $5.3 \times 10^{-7}$, $8.7 \times 10^{-7}$ | $7 \times 10^{-8}$, $1 \times 10^{-7}$ |
| Damage efficiency, $\eta$ | 0.15 | 0.45, 0.35, 0.25 | 0.4, 0.4 |
| Dimer interstitial fraction in cascade, $f_{i2}$ | 0.5 | 0.1 | 0.4, 0.1 |
| Trimer interstitial fraction in cascade, $f_{i3}$ | 0.2 | 0.1 | 0.1, 0.1 |
| Four-interstitial fraction in cascade, $f_{i4}$ | 0.06 | 0.7 | 0.06, 0.7 |
| Dimer vacancy fraction in cascade, $f_{v2}$ | 0.06 | 0.1 | 0.06 |
| Trimer vacancy fraction in cascade, $f_{v3}$ | 0.03 | 0.7 | 0.03 |
| Four-vacancy fraction in cascade, $f_{v4}$ | 0.02 | 0.1 | 0.02 |

## 3. Results

### 3.1 Cluster dynamics results and predictions for dislocation density and vacancy concentration

In this work we consider four different irradiation conditions.

1. Conditions set 1: Fast reactors at low temperature ($9.4 \times 10^{-7}$ dpa/s and 320 °C).
2. Conditions set 2: Fast reactors at intermediate temperature ($5.3 \times 10^{-7}$ dpa/s and 380 °C, average conditions of Phénix in Table 2).
3. Conditions set 3: LWR conditions at low temperature ($7 \times 10^{-8}$ dpa/s and 275 °C).
4. Conditions set 4: LWR conditions at intermediate temperature ($1 \times 10^{-7}$ dpa/s and 343 °C).

We model the first set of conditions to validate the CD model for point defects and loops as there are several consistent experimental data sets on loop behavior for these conditions. We model the second set of conditions (Sec. 3.2.3 and 3.2.4) to explore the ability to predict precipitation because of the availability of experimental data on precipitation at these and similar conditions. Finally, we model the third and fourth sets of conditions (Sec. 3.3) to gain insight into austenitic steels phase instability and precipitate evolution under the conditions of most importance for practical applications, which are for LWRs under extended life conditions. We note that these conditions were selected based on the availability of experimental data for comparison and not to explore the effects of different parameters.

*3.1.1   Conditions set 1:  cluster dynamics results for fast reactors at low temperature*

By solving governing equations in section 2.1 along with parameters in section 2.2 we are able to capture the loop evolution in 316 SS under neutron irradiation. We fit the model with three sets of experimental data, which are chosen for the following reasons. First, these experiments are all conducted at 320 °C under neutron irradiation, making them highly relevant for LWR conditions.  Furthermore, these studies go to quite high doses (up to 40 dpa), so large dose effects can be captured.  Finally, these experiments also were all conducted at the BOR-60 reactor, so the data are expected to be more consistent than samples collected from more varied environments. Figure 1 shows the evolution of loop size and number density in 316 SS under neutron irradiation at 320 °C for these three experimental data sets, along with our modeling result.

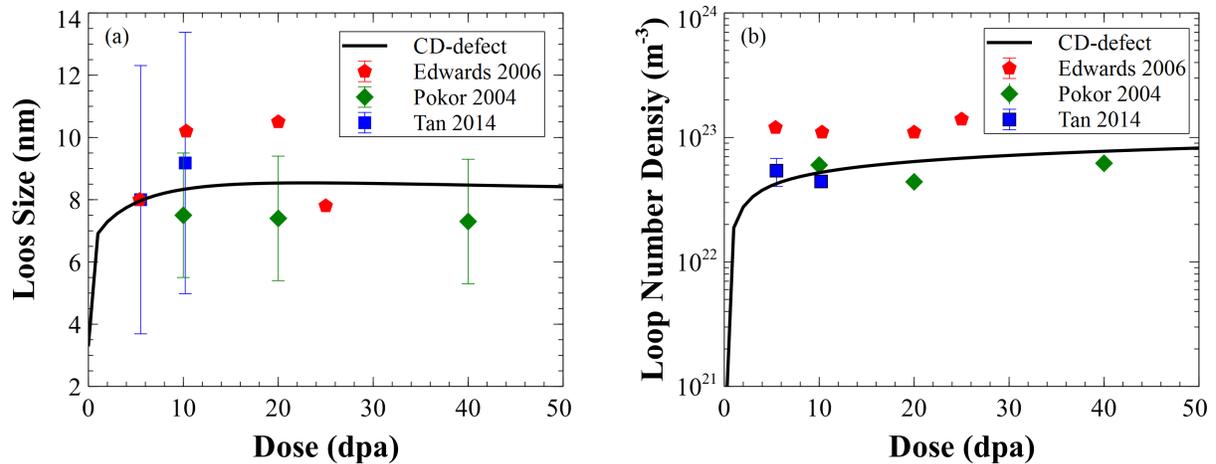

**Figure 1.** Loop size (a) and loop number density (b) evolution in 316 SS at 320 °C under 9.4×10$^{-7}$ dpa/s neutron irradiation (conditions set 1, see text).

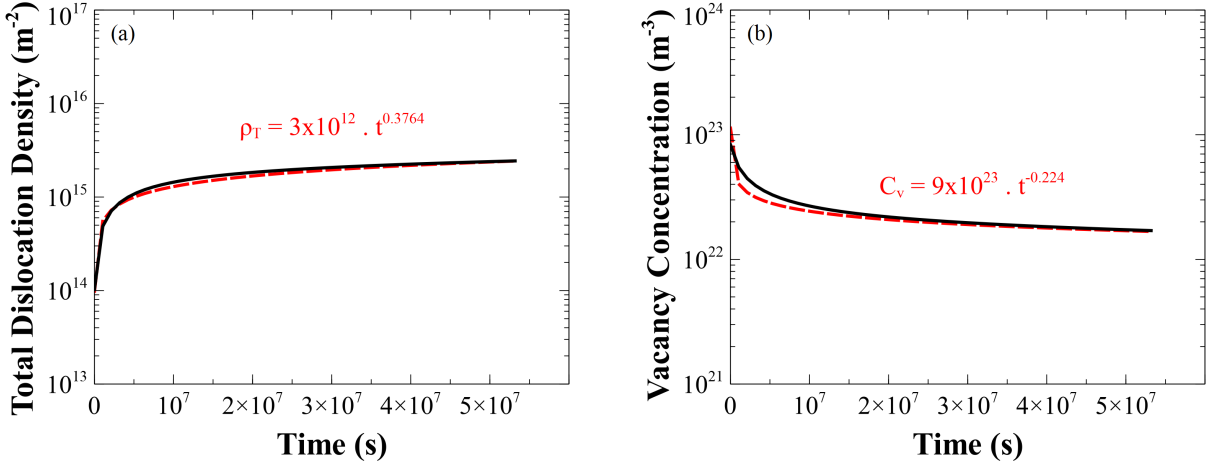

**Figure 2.** Evolution of total dislocation density (a) and single vacancies concentration (b) in 316 SS at 320 °C under 9.4×10$^{-7}$ dpa/s neutron irradiation (conditions set 1, see text). Note that the fitted lines are dashed.

The CD modeling of defect evolution uses $E_{b2i}$ as a fitting parameter, while the other materials parameters are selected from literature, as shown in Table 1. Pokor et al. [26] reported the interval of 0 to 2 eV for $E_{b2i}$ as an acceptable interval based on experimental data available in the literature. We gridded the 0 to 2 eV space and found $E_{b2i} = 0.5\ eV$ as the best value to reproduce the loop sizes and number density for conditions set 1 (9.4 × 10$^{-7}$ dpa/s and 320 °C) in Figure 1.

After finding the appropriate parameters for the CD model to reproduce the loop size and number density evolution, we use the CD model to predict the evolution of single vacancy and total dislocation concentrations, as these are critical for modeling precipitation. The single vacancy and total dislocation concentrations were then fit with simple functional forms, which were incorporated into the precipitation model as input data. If we assume the loops are circular we can calculate the total dislocation density as following,

$$\rho_{Total} = \rho + \pi D_l N_l, \tag{25}$$

where $\rho$ is the network dislocation density, $D_l$ is loop mean diameter, and $N_l$ is loop mean number density.

Extra single vacancies generated during irradiation will enhance the diffusion of substitutional elements as following [39],

$$D_X^{irr} \approx [D_V C_{1V}^{irr}]\frac{D_X^{th}}{D_{sd}^{th}} + D_X^{th} \approx (\frac{C_{1V}^{irr}}{C_{1V}^{th}}+1)D_X^{th} \approx \frac{C_{1V}^{irr}}{C_{1V}^{th}} \times D_X^{th}, \tag{26}$$

where $D_X^{irr}$ ($D_X^{th}$) is the radiation enhanced (thermal) diffusion coefficient of element $X$, $D_V$ is the diffusion coefficient of vacancy, $C_{1V}^{irr}$ ($C_{1V}^{th}$) is the concentration of single vacancies under irradiation (thermal equilibrium) and $D_{sd}^{th}$ is the self-diffusion coefficient. In addition, irradiation induced faulted loops increase the nucleation sites for those phases that nucleate at dislocations. **Figure 2** shows the evolution of single vacancy and total dislocation concentrations and their curve fitted functions versus time (in seconds). We note that in adopting these curves for use in our simulations we are assuming that the CD model results can be applied in the complete alloy simulation. In particular, this assumes that the precipitates do not significantly impact the defect concentrations. This is a reasonable approximation given the generally low precipitate concentrations in these alloys [9, 10, 13]. For example Edwards et al. [9] reported no precipitation for CW 316 SS up to 20 dpa (9.4 ×10$^{-7}$ dpa/s and 320 °C), and therefore their sink strength is zero, while at 20 dpa the sink strength of Frank loops is ~35 ×10$^{14}$ m$^{-2}$).

Our CD model does not have He and therefore, we do not expect that current model be able to capture vacancy clustering and cavity evolution correctly. However, this drawback has minor effect at the temperature range of our interest (~ 300 °C) because most experimental observations at this temperature range did not report cavity formation in CW 316 SS [9, 26, 40].

*3.1.2 Conditions set 2: cluster dynamics results for fast reactors at intermediate temperature*

After constructing the CD model we change the environment conditions (dose rate, temperature and in-cascade clustering) to the second conditions (5.3 × 10$^{-7}$ dpa/s and 380 °C) because of availability of precipitation experimental data at these conditions for model validation. Figure 3 shows the CD model results for loop size and number density and Figure 4 shows the total dislocation and vacancy concentration evolution for conditions set 2.

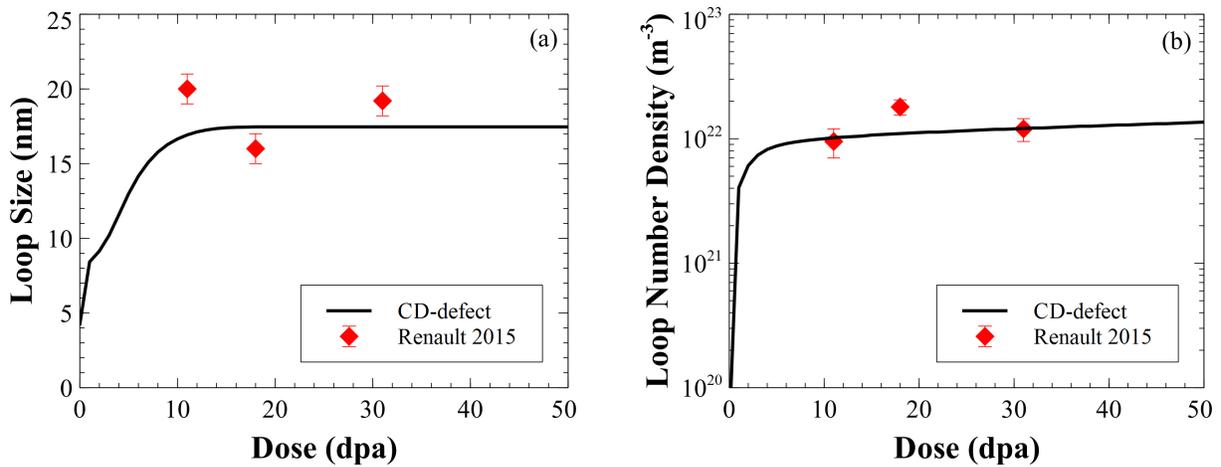

**Figure 3.** Loop size (a) and loop number density (b) evolution in 316 SS at 380 °C under 5.3×10$^{-7}$

dpa/s neutron irradiation (experimental data are from Ref. [11]) (conditions set 2, see text).

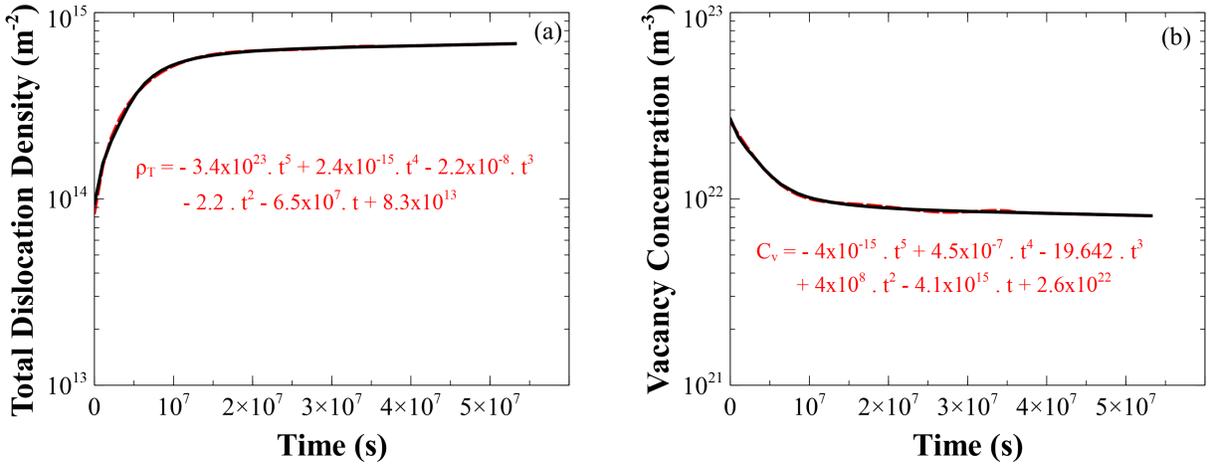

**Figure 4.** Evolution of dislocation density (a) and single vacancies concentration (b) in 316 SS at 380 °C under 5.3 ×10$^{-7}$ dpa/s neutron irradiation (experimental data are from Ref. [11]) (conditions set 2, see text). Note that the fitted lines are dashed.

### 3.1.3 Conditions set 3 and 4: cluster dynamics results for LWR conditions

To gain insight into the precipitation evolution at LWR conditions, specifically under extended life conditions, we need the CD results for these conditions up to high doses (~100 dpa). The developed CD model for fast reactors at low temperature with proper irradiation parameters (Table 2) is fairly successful in predicting the evolution of loop size and loop number density for the LWR at low temperature (conditions set 3) (Figure 5 and Figure 6).

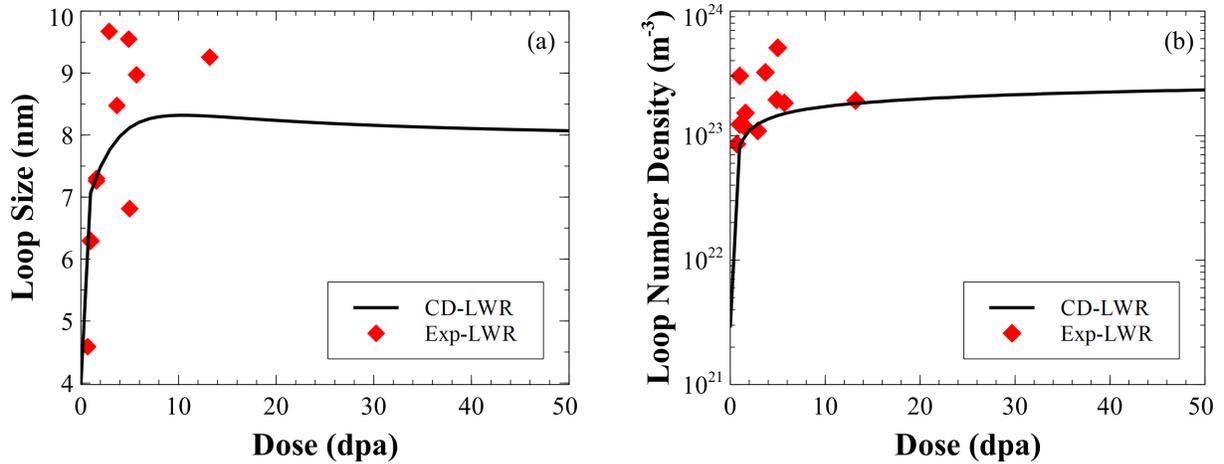

**Figure 5.** The loop size (a) and loop number density (b) evolution under LWR conditions (275 °C and 7 × 10$^{-8}$ dpa/s) compared with experimental results [37] (conditions set 3, see text).

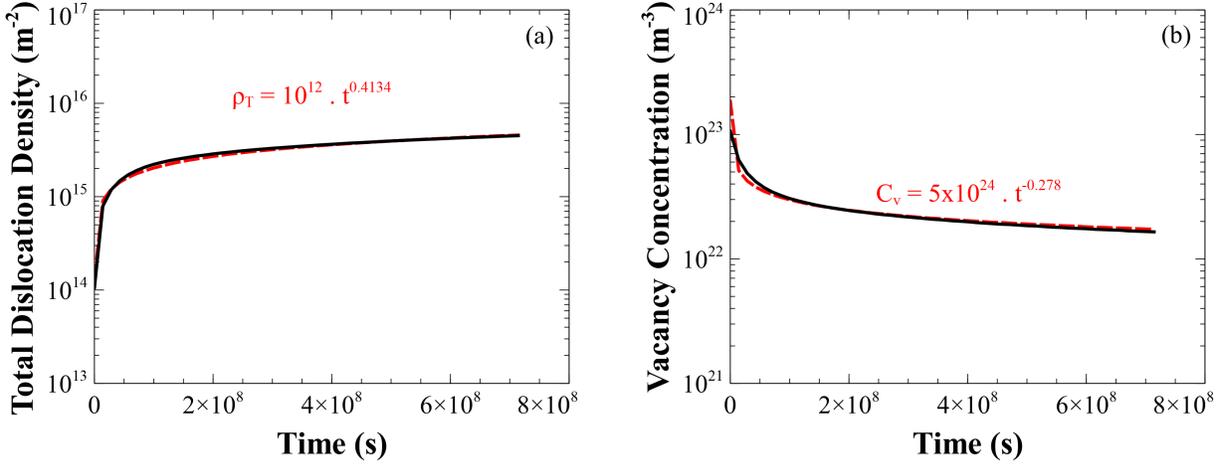

**Figure 6.** The evolution of total dislocation density (a) and single vacancy concentration (b) under LWR irradiation condition (275 °C and $7 \times 10^{-8}$ dpa/s) (conditions set 3, see text). Note that the fitted lines are dashed.

We did not provide figures for the CD results for the conditions of LWR at intermediate temperature (conditions set 4, $1 \times 10^{-7}$ dpa/s and 343 °C) because we only had one experimental datum for these conditions (average loop size = 9.5 nm and number density = $0.85 \times 10^{23}$ m$^{-3}$ at 12.2 dpa) [10].

### 3.2   Precipitation in 316 stainless steels under irradiation

#### 3.2.1   Thermodynamics

To capture the thermodynamics of 316 SS we use the OCTANT database [14, 41, 42]. OCTANT includes Fe, C, Cr, Ni, Mn, Mo, and Si with a focus on thermodynamic modeling of AISI 316 austenitic stainless steels. In this work we use the same material composition as used in reference [11]. The chemical composition of the alloy we will model is listed in Table 3. Minor alloying elements like P, S, Cu, Al, B, Nb, and Ti were not considered in this work.

**Table 3.** Chemical composition (wt.%) of 316 SS [11] used for precipitate modeling in this study.

| Alloy | Fe | Cr | Ni | Mn | Si | C | Mo |
|---|---|---|---|---|---|---|---|
| 316 SS | Bal. | 16.6 | 10.6 | 1.12 | 0.68 | 0.054 | 2.25 |

For the composition in Table 3 OCTANT predicts α ferrite at low temperatures [14]. However under routine processing conditions of austenite steels, the presence of α ferrite in the microstructure is rare, presumably due to kinetic limitations. Therefore, we suspend α ferrite in the phase calculations. Figure 7 shows the calculated equilibrium phases and their mole fractions, without α ferrite, in 316 SS from 250 °C to 400 °C.

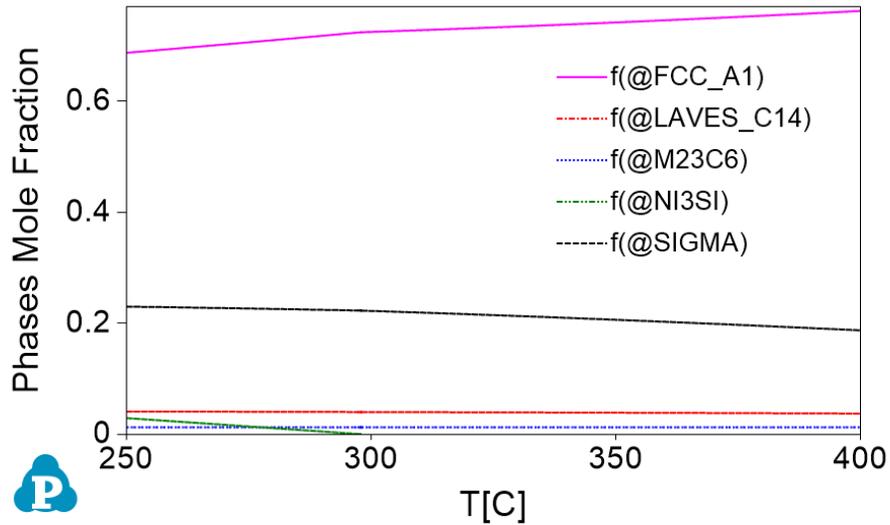

**Figure 7.** Calculated equilibrium phase mole fraction vs temperature (°C) for 316 SS.

The phase fraction study (Figure 7) shows that FCC_A1 (austenite), Sigma, Laves, and $M_{23}C_6$ are stable phases at temperatures around 300 °C. However, reported experimental data on precipitates in 316 SS under irradiation in the LWR temperature range (~ 300 °C) do not show Sigma and Laves phases [9, 11, 12, 43-45]. This fact suggests that Sigma and Laves are kinetically inhibited phases in 316 SS under LWR conditions. We note that the low temperature data is exclusively from CALPHAD predictions and no direct experimental validation has been done for these thermodynamic predictions at temperatures lower than 400 °C.

*3.2.2  Kinetics*

We combine the thermodynamics from Sec. 3.2.1 with the radiation defects modeling from Sec. 3.1 to determine the precipitation behavior in 316 SS under irradiation. For the kinetics part of the study we use the thermo-kinetic software package MatCalc developed by Kozeschnik et al. [15]. MatCalc treats the kinetics of microstructural processes based on classical nucleation theory and evolution equations for the radius and composition of each precipitate derived from the thermodynamic extremum principle [18].

Precipitation simulation needs four sets of data as input; 1) thermodynamic property database, 2) mobility database, 3) interfacial energies, and 4) microstructure information (specifically, precipitate nucleation sites). For the thermodynamic database we used OCTANT and for the mobility database we used MatCalc mobility database (mc_fe_v2.006) for steel [46], which contains elemental mobility data for face-centered-cubic (FCC) and body-centered-cubic (BCC) steels. Interfacial energies for the interface of precipitate and matrix depend on degree of coherency, crystallographic misorientation, elastic misfit strains, and solute segregation. Because of this complexity, interfacial energy generally cannot be determined reliably from direct experiment or atomistic calculations and it is usually treated as a fitting parameter in precipitation modeling. MatCalc uses a generalized nearest-neighbor broken bond (GNNBB) model for calculation an estimate of the interfacial energy [47] and we use these default values for our present, mostly qualitative, models. Finally, the microstructure information, specifically the initial dislocation density and grain size depend on heat treatment history of individual alloys and we use the values from Pokor et al. [26] which are given in Table 1.

To incorporate the effect of irradiation on precipitation we modify the MatCalc input parameters from those for thermal aging modeling to include CD-predicted RED and dislocation density, which we briefly discuss here. The RED is included by adding in the effects of excess vacancies from radiation on elemental diffusion. We assume the elemental diffusion is dominated by vacancies. Therefore, the diffusion enhancement factor can be calculated by the functions in Figure 2(b), Figure 4(b), and Figure 6(b) and applying equation (26). The challenge of including dislocation density is that generally, in precipitation under aging, the background dislocation is assumed to be stationary, which means the number of nucleation sites is unchanging during precipitation. This assumption is not correct for materials under irradiation as the Frank loops, induced by irradiation, are evolving and consequently nucleation sites are changing with time. Thus, the functions in Figure 2(a), Figure 4(a), and Figure 6(a) are used to define time dependent matrix dislocation properties in MatCalc.

Experimental observations reported the formation of carbide phases (including $M_{23}C_6$ and $M_6C$), γ' and G-phase in standard 316 SS under LWR conditions [4, 9-13, 48]. Among these phases, carbides are the most common second phases in stainless steels and they form easily under aging [14]. Unlike carbides, γ' and G-phase have not been observed in standard 316 stainless steels under thermal aging. These phases are generally Ni-Si rich clusters and are believed to form because of radiation induced segregation of Ni and Si to sinks (dislocations, grain boundaries, voids surfaces etc.). In addition to these common phases some studies showed that ferrite phase also may form under irradiation [49, 50] but due to scarcity of experimental data we exclude them from current work. In the sections below we first discuss the precipitation of carbides and then the γ' and G-phase.

*3.2.3   Carbides*

Formation of carbides in stainless steels is well established. $M_{23}C_6$ is one of the first precipitate phases that show up in 316 SS under aging [51] and its thermodynamics has been extensively studied at high temperatures [51, 52].

A question of some interest to the community is how we expect radiation to enhance carbide kinetics, if at all. Excess vacancies introduced by irradiation enhance the diffusion of substitutional elements and have minor effect on interstitial atoms. In $M_{23}C_6$ the M sublattice includes substitutional elements (Cr, Fe, Mn, Mo, Ni) but carbon comes from interstitial sites. Therefore, radiation induced excess vacancies only enhance the diffusion of M sublattice elements in $M_{23}C_6$. Diffusion coefficient databases show that carbon is the fastest diffusing element in 316 SS under thermal aging [53] and consequently the kinetics of precipitation of carbide phases is controlled by M elements, which are the slowest diffusers in the compound. Under irradiation the mobility of M elements increases dramatically, which means that carbides formation kinetics is expected to be enhanced by irradiation. However, it is important to assess whether carbon is still the fastest diffusing element in the carbides under irradiation, as the extent of the radiation enhancement will be reduced if carbon becomes the limiting element, since it is not impacted significantly by the irradiation. To estimate the enhancement of M elements we note that the concentration of excess single vacancies under irradiation converges to $\sim 10^{22}$ m$^{-3}$ based on Figure 2. At 320 °C the thermal concentration of single vacancies is about $10^{16}$ m$^{-3}$ ( $C_{1V}^{eq} = \exp(S_{fv}/k)\exp(-E_{fv}/kT)$ where $S_{fv}$ is the vacancy formation entropy, $E_{fv}$ is vacancy formation energy, $k$ is Boltzmann constant, and $T$ is temperature – see Table 1 for values). Comparing the vacancy concentration under irradiation and thermal aging shows that irradiation will enhance the diffusion coefficients of substitutional elements by a factor of $\sim 10^{22}/10^{16}=10^6$. Table 4 shows the thermal and irradiation enhanced diffusion coefficient of carbon and M elements (Cr, Fe, Mn, Mo, Ni) in 316 SS at 320 °C. Diffusion coefficient data under irradiation show that the radiation enhanced diffusivity of M elements are still less than carbon, which indicates that M elements will still control the kinetics of carbide formation under irradiation at 320 °C and 9.4×10$^{-7}$ dpa/s. For LWR conditions the neutron flux is lower than 9.4×10$^{-7}$ dpa/s, which is expected to even further reduce M diffusion versus carbon. Therefore, the carbide formation kinetics is expected to be enhanced by radiation, and the enhancement is expected to arise from the enhancement of the transport of the metal atoms, without additional limiting factors associated with carbon.

**Table 4.** Diffusion coefficient of carbide components in 316 SS at 320 °C (from MatCalc mobility database, mc_fe_v2.006 [46]).

| Element | Tracer diffusion coefficient at FCC (m$^{-2}$/s) | Radiation enhanced diffusion (m$^{-2}$/s) |
|---|---|---|
| C | 6e-20 | 6e-20 |
| Cr | 1e-29 | 1e-23 |
| Fe | 9e-30 | 9e-24 |
| Mn | 8e-29 | 8e-23 |
| Mo | 2e-29 | 2e-23 |
| Ni | 2e-30 | 2e-24 |

By incorporating enhanced diffusion of M atoms and dislocation evolution (which alters heterogeneous nucleation sites) in our thermo-kinetic model, we are able to model the carbides precipitation under irradiation. Since we have better experimental data on carbides precipitation at temperature around 390 °C [11, 12, 43, 45] we focus on this temperature range. We get CD input data for this condition form Sec. 3.1.2.

Figure 8 shows both modeling and experimental data of carbides volume fraction in 316 SS at a range of temperatures lower that 400 °C under radiation ($8.7 \times 10^{-7}$ dpa/s). Most experimental data show very low carbide volume fraction even up to 100 dpa. In addition to the reported data in Figure 8, Tan et al. [13] and Edwards et al. [9] also reported very small nanoscale carbide precipitates under irradiation at BOR-60 at 320 °C, which further supports the result that there is very little carbide precipitation under irradiation.

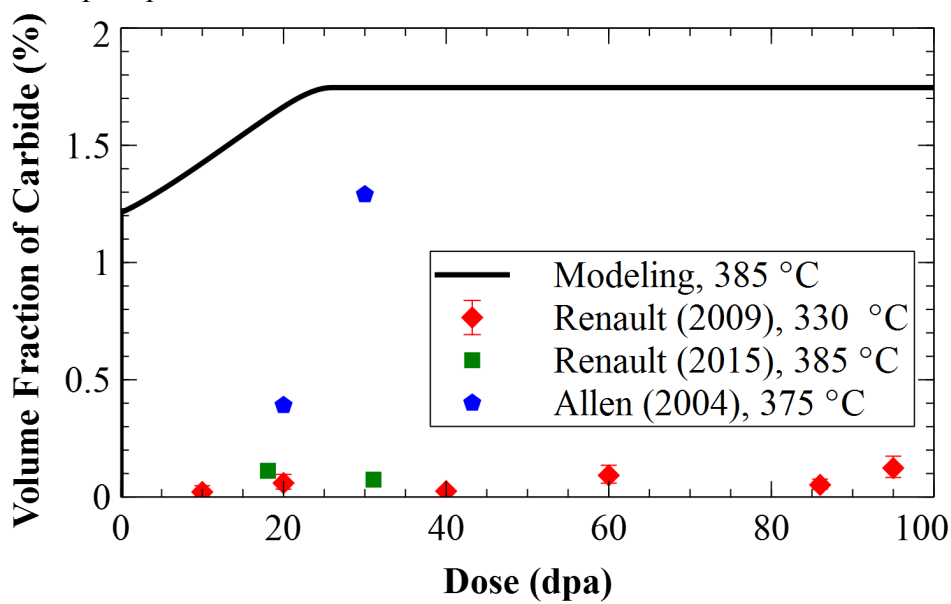

**Figure 8.** Evolution of total carbides ($M_{23}C_6 + M_6C$) volume fraction compared to experimental data [11, 12, 45] (modeling dose rate = $8.7 \times 10^{-7}$ dpa/s). We note that the nature of precipitates in Renault (2009) were uncertain and could be carbide or G-phase.

Comparing the volume fraction of carbides predicted by CD-informed thermo-kinetic model with experimental data generally shows an approximately order of magnitude overestimate from the model. Modeling under the conditions of Figure 8 predicts a relatively rapid formation of a 1.2% volume fraction $M_{23}C_6$ (this phase saturates within ~0.1 dpa), which then gets replaced gradually by $M_6C$ while increasing to a final volume fraction of 1.7%. Formation of $M_6C$ has been reported at high temperature (650 °C) aging of 316 SS [51]. Goldschmidt also reported the formation of $M_6C$ from $M_{23}C_6$ which is consistent with our prediction [54].

Despite the high volume fraction of carbide in modeling, the bulk of experimental data show a steady state volume fraction in the order of 0.1%. The exception is the data of Allen et al. [45], particularly the value at ~30 dpa, which shows a volume fraction of about 1.3% (all precipitate formation percentages are given as volume fractions in this work unless otherwise labeled). While we cannot presently be sure of the explanation of the differences in this data vs. other experiments, it is observed that the highest value is comparable to the values predicted by our model.

We note that for carbide precipitation we use the initial 316 SS bulk composition and ignore the effect of RIS on bulk composition. This approximate is reasonable because the carbide forms very quickly on the timescale of the RIS, which takes tens of dpa to saturate. However, one effect of RIS on carbide could be a shift in $M_{23}C_6$ to $M_6C$ transformation. Aging studies suggest that carbide forms on dislocations and grain boundaries [51]. After carbide formation, the composition at these dislocations and grain boundaries would gradually change due to RIS. Specifically, RIS would deplete the Mo at sinks (Table 5) which is believed to facilitate the $M_{23}C_6$ to $M_6C$ transformation [51, 54]. Therefore, the $M_{23}C_6$ to $M_6C$ transformation might become slower due to Mo depletion through RIS.

To determine the source of the discrepancy between the model and experiments we first considered possible errors in the model thermodynamics. We benchmarked our thermodynamic database against aging experimental data at high temperatures where kinetics allows the system to reach equilibrium. The key question we seek to answer is whether the carbide saturation level of volume fraction (1.2-1.7%) is reasonable or not. It is well known that $M_{23}C_6$ is the first emerging second phase in 316 SS during aging and it reaches to its saturation level fairly fast [51]. The saturation level of carbides is also known experimentally to vary between 1-2 wt.% depending on initial carbon concentration [52]. Given that our irradiation experiments are at lower temperature than the experiments in Ref. [52] (this reference has several aging experiments in the range of 600-700 °C) we expect similar or higher concentrations of carbides at equilibrium. Therefore, our rapid formation of 1.2% carbide followed by additional growth to 1.7% is not surprising, and in fact is quite consistent with what is known about carbide precipitation thermodynamics. Therefore, from a thermodynamics perspective, it is the low volume fraction of carbides in the irradiation experiments that are surprising.

We have thought of four possible mechanisms that may cause the discrepancy we observe between the experimental and modeling results, and these mechanisms are

1. In the experimental period the carbide does not reach equilibrium due to slow kinetics.
2. Formation of other phases, e.g. γ' and G-phase, suppress the carbide.

3. Irradiation changes on spatial distribution of carbon alter or suppress carbide nucleation and growth, leading to less carbide and/or harder to detect small carbides.
4. Carbide particles are dissolved by irradiation.

The first mechanism is unlikely as experiments go up to very high doses (e.g., 100 dpa), so it is hard to believe that a radiation enhanced phase, like the carbides, does not reach to its equilibrium under the experiments. The second mechanism is also unlikely as our kinetics simulation with RIS composition, where the $\gamma'$ and G-phase can form, shows formation of ~ 1.2% carbide. Shim et al. [16] also reported the formation of carbide in RIS regions [16], suggesting the RIS induced phases do not suppress the carbide.

We believe the third and/or fourth hypothesis could be the causes behind low experimental and high modeling volume fraction of carbide. In support of the third mechanism, Jiao and Was [55] reported that carbon may deplete at grain boundaries under proton irradiation. In another work, Hatakeyama et al. [56] studied the elements segregation at dislocations and their results showed that the carbon level at dislocations is similar to the matrix in the irradiated 316 SS (same results were reported by Jiao and Was for grain boundary). Unlike Jiao and Was [55], who compared to an unirradiated carbon profile, Hatakeyama et al. did not have an unirradiated profile showing carbon segregation at dislocations. However, based on the observation of Jiao and Was [1], we expect that the segregation behavior of elements at dislocation and grain boundary is similar. Therefore, it is likely that C segregates at dislocation before irradiation (similar to grain boundary) and Hatakeyama's results of non-segregated C at dislocation in irradiated 316 SS suggest C is depleted under irradiation at the dislocation.

On the other hand, thermal aging experimental studies showed that carbides prefer to nucleate at grain boundaries and dislocations [51]. Therefore, we can conclude that since irradiation pushes carbon away from the preferred sites for carbides, e.g. dislocations and grain boundaries, the ability to nucleate new carbides or grow existing ones may be suppressed under irradiation. This suppression may reduce the total carbide precipitation and/or shift carbides to nucleate more homogeneously, which will likely lead to the formation of smaller precipitate particles that are harder to detect experimentally.

Radiation enhanced dissolution, the fourth mechanism above, also may be the cause of discrepancy between the model and experimental results. Similar to modeling results, experimental results show that carbide saturates at the very beginning of irradiation. This fact indicates that carbide phases under irradiation at fairly low temperatures ranges (e.g. LWR temperature range) will quickly reach a steady state, consistent with our prediction of rapid kinetics for the carbides. However, this steady state does not appear to be thermodynamic equilibrium. To understand this steady state we note that irradiation qualitatively affects the phase formation and stability in two opposite ways; 1) it enhances the precipitation by increasing the diffusion of elements in matrix; 2) it dissolves the precipitates through recoil dissolution and disordering dissolution [57]. Competition between these two effects provides a steady state situation for precipitate phases that is not thermal equilibrium. Enhancing and dissolving effects of irradiation guide precipitate particles to an optimum size. Precipitates smaller than the optimum size will grow and precipitates larger than this size will shrink. The latter effect is sometimes called radiation induced reverse Ostwald ripening. We propose the radiation enhanced dissolution as one possible reason that carbides are less than thermal equilibrium at fast reactors conditions of 300-400 °C. In other words, we propose the dissolution effects, which are not presently in our model, are the reason for the above discussed discrepancies in predicted (and measured at high-temperature) thermal equilibrium carbide volume fraction and low-temperature radiation induced carbide volume fraction. While further work is needed to verify this hypothesis we note that carbides dissolution under irradiation was experimentally observed in PE16 alloys [58]. The role of this dissolution may depend strongly on many factors (composition, temperature, etc.) but particularly flux, making it a critical area for further research given the use of high-flux accelerated testing methods in many studies of microstructure evolution.

The mechanism of radiation enhanced dissolution may help us to understand the Allen et al. [45] data points. The dose rate in Allen's experiment is $1\times10^{-7}$ dpa/s while the Renault's experiments were conducted at $5.3\times10^{-7}$, and $8.7\times10^{-7}$ dpa/s. All these three experiments were done on CW 316 SS and although Allen et al. did not mention the initial concentration of carbon we know that the typical 316 SS have maximum 0.1 wt.% carbon [52]. Renault's 316 SS had 0.06 wt.% carbon. Based on available aging experimental data [52], it is unlikely that 0.04 wt.% difference in carbon concentration could cause 12 times higher carbide precipitate, i.e. the values seen for 30 dpa in Allen et al.. The aging experimental data show that 0.05 wt.% higher carbon can increase the carbide precipitate level by about three times (from 0.7wt.% to 2wt.%) [52]. In addition, the experiments of Allen et al. differ by just 10 °C from Renault et al., which suggests temperature difference is not expected to have a large effect on the total steady state carbide volume fraction. Therefore, even if we assume that Allen's alloy had the highest likely initial carbon, its carbide level is not expected to be 12 times higher that Renault's alloy. The most obvious difference between these experiments that might explain the different volume fractions is that Renault's dose rate for the data point at 30 dpa is $8.7\times10^{-7}$ dpa/s, while Allen's dose rate is $1\times10^{-7}$ dpa/s. This difference suggests that the lower dose rate experiment of Allen leads to weaker radiation enhanced dissolution than Renault, and therefore larger precipitate volume fractions that appear to take longer to reach steady state. However, it is also possible that Allen's samples were exceptionally high in initial matrix carbon, which led to a very high carbide precipitate volume fraction.

*3.2.4 γ′ and G-phase*

γ′ is an ordered cubic phase ($L1_2$, $F_{m3m}$) with almost similar lattice parameter as austenite and little or no misfit. The γ′ stoichiometric atomic composition is $Ni_3X$ where X typically is Si, Nb, or Al [7]. Morphologically, γ′ was observed as small spheres in the matrix [59].

G-phase also has a cubic crystal structure ($A1$, $F_{m3m}$). It is a complex silicide with stoichiometric atomic composition of $M_6Ni_{16}Si_7$ where M can be Cr, Fe, Mn, Mo, and Ti. G-phase has 1.1 nm lattice parameter and its unit cell contains 116 atoms. Morphologically, G-phase was observed as small rods [59].

Aging studies of 316 SS have shown that γ′ and G-phase are not thermally stable phases in standard 316 SS at temperatures where they are often seen under irradiation, so they are categorized as radiation induced phases [60]. The main driving force for formation of γ′ and G-phase is radiation induced segregation (RIS). In typical 316 SS the level of Si is low enough (~0.7 wt.%) to prevent formation of γ′ and G-phase. However, under irradiation the Si and Ni enrich at sinks (e.g. dislocations) and these excess Si and Ni in RIS regions facilitate the formation of γ′ and G-phase. Segregated elements at sinks change the alloy composition in these regions to such an extent that those regions can be approximately considered as different alloys. In fact, in some literature the term "microalloys" is used for RIS regions [5] and we will adopt this nomenclature. Table 5 shows the typical composition of RIS enhanced microalloy regions in 316 SS at 5 dpa, 10 dpa, and 20 dpa.

**Table 5.** Typical composition of radiation induced segregation regions for 316 stainless steels (wt.%) [16].

|  | Fe | Cr | Ni | Mo | Mn | Si | C |
|---|---|---|---|---|---|---|---|
| RIS composition at 5 dpa | Bal. | 14±2 | 18±2 | 1 | 1 | 3±2 | 0.05 |
| RIS composition at 10 dpa | Bal. | 12±1 | 21±4 | 1 | 1 | 5±1.5 | 0.05 |
| RIS composition at 20 dpa | Bal. | 11±2 | 24±2 | 1 | 1 | 6 | 0.05 |

In this study we focus on precipitation behavior of CW 316 SS as they have been studied extensively experimentally. In experimental observations it is generally reported that the γ′ is the dominant phase in CW 316 SS and G-phase is almost suppressed [9-11, 13, 44, 45, 61, 62]. In fact, among the eight experimental works that we have found for CW316, covering 10 to 90 dpa, seven of them do not report G-phase, and although Renault, et al. [11] showed some G-phase, it was very little (0.04% volume fraction). These results suggest that the stability of G-phase under the RIS composition is highly uncertain, and that our CALPHAD thermodynamic model is perhaps overstabilizing the G-phase. It is also possible that our interfacial energy for G-phase is too small, and therefore its nucleation kinetics is too fast, and we are nucleating G-phase under circumstances where it has still not formed in experiments. As noted above in Sec. 3.2.2, our interfacial energies are determined by a simple method which one cannot expect to be quantitative. Given the limited data, further refinement of the G-phase free energy and/or interfacial energy is quite challenging, and beyond the scope of the present study. Therefore, we simply suspend the G-phase in our modeling of the kinetics of CW 316 SS. However, we note that the tools we have developed here can easily be applied to other situations where G-phase might form, although a refinement of the interfacial energy, likely to larger values that we have used, may be needed. Transmission electron microscopy (TEM) observations also show that $\gamma'$ nucleates at Frank loops [7] and it is well distributed in the matrix. Since the radiation induced segregation is necessary to form $\gamma'$, we can conclude that $\gamma'$ forms in the matrix of microalloys, i.e., the RIS region around dislocations. Note that we do not consider the RIS and microalloys around grain boundaries as their contribution to the sink density is negligible compared to that from the dislocations.

The ability to model an alloy with a RIS microalloy environment around dislocations is not presently available in the MatCalc code that we are using to model precipitation (see Sec. 3.2.2) and recoding to rigorously treat this situation is beyond the scope of this work. Therefore, we have taken an approximate approach to model the precipitation, which is to simulate the precipitation in the RIS region by using just the RIS microalloy environment in MatCalc. In other words, to study the RIS region we study a bulk material with the RIS composition and an appropriate dislocation density and RED. A similar approach was taken by Shim et al. [16] for modeling radiation induced phases in 316 SS. However, we extend their approach by including the loop density and radiation enhanced diffusion evolution from the models we have developed (see Sec. 3), which enables us to capture the $\gamma'$ more quantitatively.

There are at least two major approximations associated with our microalloy approach. First, while we can take RIS composition from experiments, it is actually a function of time. Since our model cannot capture the continuous change of RIS microalloy composition we assume that the composition is fixed at each dpa, with values taken from experiments (or interpolation between the experiments). This assumption will not affect the results significantly as the kinetics of $\gamma'$ is very fast compared with RIS evolution time scale, so we can approximately take RIS as fixed for any given simulation of $\gamma'$ evolution. Figure 9 shows the kinetics of $\gamma'$ precipitation inside the RIS region, which indicates that the $\gamma'$ reaches to its equilibrium level within 0.01 dpa. This value is much lower than that needed for any major compositional change due to RIS, which typically takes tens of dpa to saturate.

The second approximation comes from the assumption that the γ' formation is shut down by the depletion of local enhanced concentrations (primarily Si). This approximation is again reasonable due to time scales. Because the γ' formation takes place over ~0.01 dpa and is driven by the RIS enhanced microalloy composition which evolves over tens of dpa, it is expected that the γ' will precipitate until the local microalloy supersaturation is depleted.

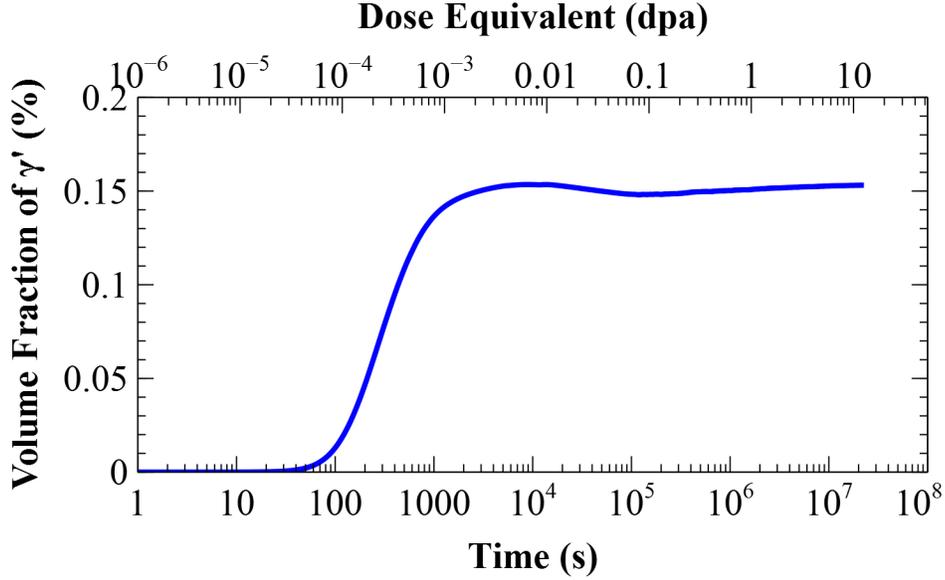

**Figure 9.** Kinetics of γ' inside the RIS region under $5.3 \times 10^{-7}$ dpa/s at 380 °C (The domain composition is the RIS composition at 5 dpa from Table 5).

Modeling the kinetics of precipitation in microalloys with the OCTANT database and MatCalc will give us the volume fraction of γ' inside the RIS region. In our calculations we consider the asymptotic volume fraction of γ' due to its fast precipitation process. To obtain the total volume fraction of γ' in austenite matrix we must multiply the obtained γ' volume fraction with the volume fraction of RIS regions (which comes from loops size and number density) as follows

$$V_{f,\gamma'}^{Aus} = V_{f,\gamma'}^{RIS} \times V_{f,RIS}^{Aus}, \tag{27}$$

here $V_{f,\gamma'}^{Aus}$ is the volume fraction of γ' in austenite, $V_{f,\gamma'}^{RIS}$ is the volume fraction of γ' in RIS regions, and $V_{f,RIS}^{Aus}$ is the volume fraction of RIS regions in an austenite matrix. $V_{f,RIS}^{Aus}$ can be calculated as following,

$$V_{f,RIS}^{Aus} = \pi D_{Loop} \times \pi r_{Cyl}^2 \times N_{Loop}, \tag{28}$$

where $D_{Loop}$ is the average diameter of loops (so $\pi D_{Loop}$ is the loop length), $r_{Cyl}$ is the radius of RIS regions width (which is ~2 nm [56]), and $N_{Loop}$ is the number density per unit volume of loops. The Frank loop data (size and number density) come from CD model.

We use the RIS compositions in Table 5 with some linear interpolation/extrapolation, when needed, and experimental loop data from reference [11] to predict the evolution of γ' in CW 316 SS under irradiation.

Using the methodology described above, we track the evolution of volume fraction of γ' under irradiation. Figure 10 shows the comparison between predicted γ' volume fraction evolution and the experimental data. The results show a good agreement between the experimental data and the integrated model (CD + CALPHAD + MatCalc) predictions. This success gives us confidence to explore the model predictions for γ' precipitation at lower flux of relevance for realistic LWR conditions.

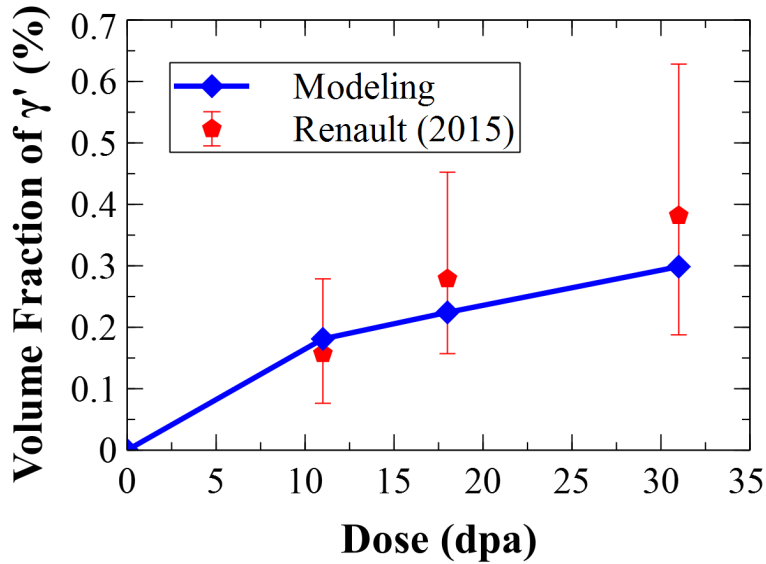

**Figure 10.** Comparison between modeling predictions and experimental data [11] of γ' volume fraction evolution in CW 316 SS under $3.5 \times 10^{-7}$, $5.3 \times 10^{-7}$, and $8.7 \times 10^{-7}$ dpa/s irradiation at 380, 381, and 386 °C, respectively. The plot is produced based on this assumption that the element segregation to dislocation is similar to grain boundaries.

*3.3 Extrapolation of the model to LWR conditions*

The integrated model (CD + CALPHAD + MatCalc) described in preceding sections was mainly developed for fast reactors where a significant range of experimental data are available. Unfortunately, the experimental database for post-irradiation microstructural examination of austenitic stainless steels under LWR conditions is insufficient to aid in verification of such model. Specifically, there is no experimental data for LWR 316 SS under extended life conditions. The developed model can therefore be used to gain insight into the less well explored domains of austenitic stainless steels degradation under LWR conditions, especially under extended life conditions. Light water reactors operate at relatively low temperature range, i.e. 275-340 °C and low dose rate $(2\text{-}11) \times 10^{-8}$ dpa/s. Fast reactors could have the same temperature range, but the dose rate is usually ten to hundred times higher [37].

The first step in using the integrated model is finding the total dislocation and single vacancy evolution under the LWR conditions. We use the CD predicted values for these quantities in Sec. 3.1.3 as input for the precipitation model.

Figure 11 shows the precipitation of carbide phases ($M_{23}C_6$ and $M_6C$) under LWR conditions. The first noticeable difference between the carbide precipitation under LWR (Figure 11) and fast reactor (Figure 8) conditions is the time lag in the transformation of $M_{23}C_6$ to $M_6C$ in LWR conditions. In fast reactors the transformation of $M_{23}C_6$ to $M_6C$ happens from the very beginning of irradiation and $M_6C$ is the dominant phase after 25 dpa (see Figure 8). However, in LWR conditions $M_{23}C_6$ to $M_6C$ transformation starts around 20 dpa and it is much slower than fast reactors, such that $M_{23}C_6$ will remain the dominant phase up to 100 dpa. We note that these dpa values are quite approximate as the interfacial energies are obtained by very approximate methods (see Sec. 3.2.2). Furthermore, the radiation induced dissolution that we hypothesize is occurring for carbides might greatly enhance this transition by enabling facile dissolution and reprecipitation. Overall, understanding that the dpa scale for significant transformation of the carbides may reside within the dpa range of LWR life extension is an important motivation for developing more accurate models for this process.

The other important factor that may cause a big difference in carbide precipitation between LWR and fast reactors is the effect of neutron flux on radiation enhanced carbon segregation and/or dissolution. As we discussed in Sec. 3.2.3 we believe that radiation enhanced carbon segregation and/or dissolution are potentially important factors in precipitation of carbide phases under irradiation. At LWR conditions the dose rate is much lower than fast reactors [37]. Therefore, it is plausible that the radiation enhanced carbon segregation and/or dissolution will be weaker under LWR conditions compared with fast reactor conditions. If that is the case, the volume fraction of carbide would be higher in austenitic stainless steels under LWR conditions compared with fast reactors at the same dose and temperature. The predicted volume fraction from the present model, which does not include radiation enhanced carbon segregation and/or dissolution, provides an upper bound for the likely volume fraction. We represent the unknown influence of these additional factors by giving a range of LWR values in Figure 11 that includes values starting from those observed in high-flux fast reactor experiments up to those predicted by our model.

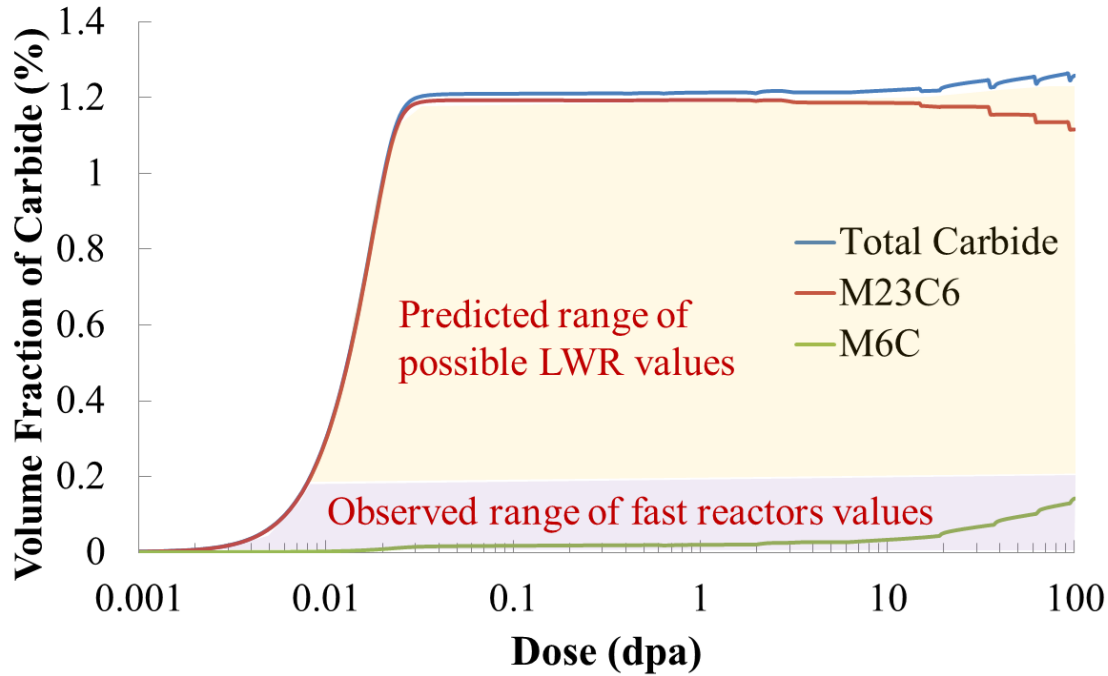

**Figure 11.** The evolution of total carbide ($M_{23}C_6 + M_6C$) volume fraction under LWR conditions (275 °C and $7 \times 10^{-8}$ dpa/s). The "Predicted range of possible LWR values" covers from those values observed under high flux fast reactor conditions to those predicted from our thermodynamic and kinetic model with no radiation enhanced dissolution, highlighted to represent. We believe the lower neutron flux in LWR conditions can cause higher carbide volume fraction compared with fast reactors due to weaker radiation enhanced carbon segregation and/or dissolution.

Edwards et al. [10] studied experimentally a CW 316 SS baffle bolt which was extracted from the Tihange pressurized water reactor (PWR). They used scanning electron microscopy (SEM) and transmission electron microscopy (TEM) for microstructural characterization and energy dispersive X-ray spectroscopy (EDS) for microchemical analysis. In their characterization on the bolt shank, which had been irradiated to 12.2 dpa at 343 °C, they reported the formation of 0.08% volume fraction of $\gamma'$ (number density of $0.6 \times 10^{23}$ m$^{-3}$ and average size of ~3 nm) and 0.64% volume fraction of an unknown phase ($0.2 \times 10^{23}$ m$^{-3}$ and average size of ~8.5 nm). They believed that the identity of the unknown precipitate phase might be some type of carbide. We used our integrated model (CD + OCTANT + MatCalc) to benchmark our prediction against Edwards' result. Edwards did not report the dose rate. Therefore, we use $1 \times 10^{-7}$ dpa/s which is slightly higher than typical LWR dose rate, e.g. $7 \times 10^{-8}$ dpa/s. These are conditions set 4 in Sec. 3.1.3. We use the CD predicted vacancy concentration to estimate the RED and then incorporate the RED and CD predicted dislocation density evolution into MatCalc. We use the Edwards' 316 SS composition as listed in Table 6.

Table 6. Chemical composition (wt.%) for Edwards' 316 SS [10].

| Alloy | Fe | Cr | Ni | Mn | Si | C | Mo |
|---|---|---|---|---|---|---|---|
| 316 SS | Bal. | 16.7 | 12.36 | 1.89 | 0.72 | 0.028 | 2.64 |

The carbon wt.% in Edwards' alloy is lower than the 316 SS that we considered in this work (Table 3). Therefore, we expect to obtain lower carbide volume fraction compared to what we predict for LWR in Figure 11. Figure 12 shows the comparison between the integrated model prediction and the Edwards' volume fraction for carbide phase. The model works very well in predicting the experimental carbide volume fraction. We note that to validate the full model we need more experimental data. However, the agreement with Edwards, et al.'s single experimental datum provides some support for our hypothesis that the carbide volume fraction under LWR conditions can reach its thermodynamic asymptotic value, which we do not usually see in fast reactors. We also note that the Edwards et al. [10] did not directly identified the observed phase as carbide, and its nature is therefore somewhat uncertain. Since the lattice parameters of $M_{23}C_6$, $MC_6$, and G-phase are similar, it is not possible to distinguish these phases without chemical analysis (APT, STEM-EDX, EFTEM, …). Therefore, it is possible that the phases that Edwards et al. reported are G-phase rather than carbide. Edwards et al. [10] quoted other works available at the time that had found carbide in their studies [44, 63] and therefore stated that this unknown phase could be carbide. Furthermore, a pair of recent results suggest that carbides are a plausible explanation for what Edwards et al. observed, although they are by no means definitive. Very recently, Fujii and Fukuya [64] have used atom probe tomography to study the CW 316 SS specimens irradiated up to 74 dpa under PWR conditions. They reported the formation of $\gamma'$ and some Ni-Si-Mn clusters which they believed could be the precursors of G-phase precipitates, but they did not report explicit G-phase even up to 74 dpa (1.5 x $10^{-7}$ dpa/s and T=305 °C). This is an argument that G-phase is not expected at Edwards, et al.'s dose of 12 dpa, supporting Edwards, et al.'s identification of their well-formed phases (not precursors) as carbides. However, Fujii and Fukuya [64] did not report the formation of carbides, which would argue against Edwards, et al.'s interpretation. Also fairly recently, Renault et al. [12] reported the formation of less than 0.05% volume fraction for G-phase in CW 316 SS at 31 dpa (8.7 x $10^{-7}$ dpa/s and T=386 °C). Comparing these results with the much larger 0.7% volume fraction observed by Edwards, et al. at a lower dose of just 12 dpa again suggests that Edwards, et al. observed carbides, not G-phase. However, we note that Edwards, et al.'s experiments are at 1 x $10^{-7}$ dpa/s (this is a typical dose rate for PWR conditions, Edwards et al. did not report the dose rate) and T= 345 °C, which are broadly similar conditions to the above studies but still significantly different in temperate and flux, making direct application of these more recent results to interpreting those from Edwards, et al. uncertain. We believe the overall preponderance of results suggests that Edwards observed carbides, but it is clear that more studies are needed to make any definitive statement. Given the limited data available under these LWR conditions we believe it is useful to include a comparison to Edwards, et al.'s data assuming it is carbides, despite the uncertainty in the phase identification.

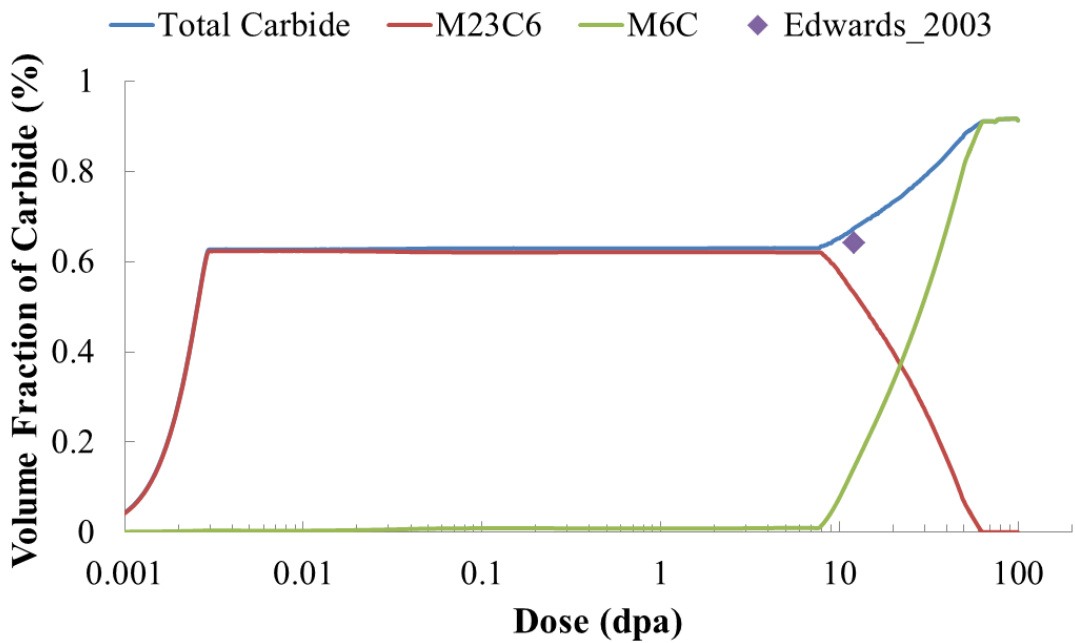

**Figure 12.** Comparison between integrated model (CD + OCTANT + MatCalc) prediction and experimental carbide volume fraction in Edwards' 316 SS under $1 \times 10^{-7}$ dpa/s irradiation at 343 °C [10]. The good agreement between modeling and experiment supports our hypothesis that the carbide volume fraction under LWR conditions might reach levels significantly higher than typically seen in fast reactors, although significantly more experimental study is needed to validate this hypothesis and our model.

For radiation induced phases, i.e. $\gamma'$ and G-phase, if we assume that the $\gamma'$ domination in CW 316 SS observed under fast reactor conditions is also true for LWR conditions, we can make some predictions about the $\gamma'$ volume fraction. First, we assume that all $\gamma'$ form at Frank loops. For $\gamma'$ precipitation on Frank loops we use the same microalloy methodology we described in Sec. 3.2.4. We assume that $\gamma'$ forms inside a cylindrical RIS region around the loops. Bruemmer et al. [65] reported the evolution of Si segregation at grain boundaries versus neutron irradiation dose up to 9 dpa for LWR conditions. The maximum Si concentration at 9 dpa is 5 wt.%. For Si concentration at higher doses we extrapolate the Si segregation linearly following the Figure 13.

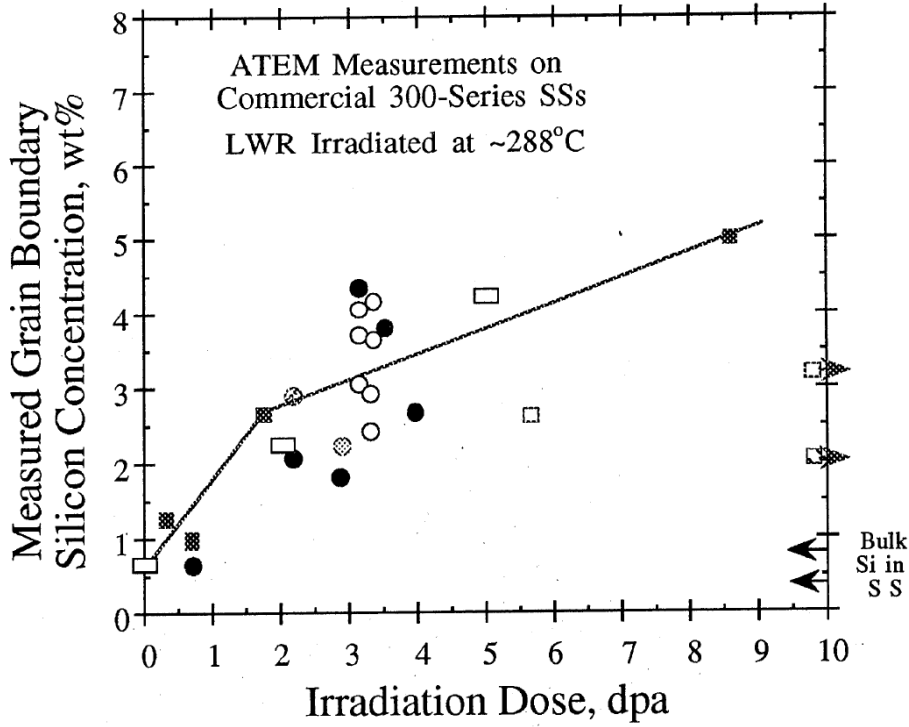

**Figure 13.** Radiation-induced grain boundary Si concentrations versus neutron irradiation dose for LWR conditions, from Ref. [65].

However, the total segregated Si is bounded by the total initial Si content of the alloy. We can find the ultimate possible Si concentration in loops based on the volume fraction of Frank loops and initial concentration of Si in the matrix as follows:

$$W_{Si}^{Aus} = W_{Si}^{RIS} \times V_{RIS}^{Aus},  \qquad (29)$$

where $W_{Si}^{Aus}$ is the weight percent of Si in matrix, $W_{Si}^{RIS}$ is the weight percent of Si in RIS region, and $V_{RIS}^{Aus}$ is the volume fraction of RIS region in matrix, which can be calculated by equation (28). For example the linear extrapolation of Figure 13 would give us 9 wt.% Si in the RIS region at 20 dpa, but the maximum Si concentration at loops possible even under the assumption that all Si in the matrix has segregated to loops, based on equation (29), would be 7.5 wt.%. ($W_{Si}^{Aus} = 0.68\,wt.\%$, initial Si,

$V_{RIS}^{Aus} = \pi D_{Loop} \times \pi r_{Cyl}^2 \times N_{Loop} = \pi(9.5 \times 10^{-9}) \times \pi(2 \times 10^{-9})^2 \times 2.4e23 = 0.09$, $D_{Loop}$ and $N_{Loop}$ come from CD model, Figure 5, and $r_{Cyl}$ is taken from Ref. [56]). If we assume that RIS increases following a linear extrapolation of Figure 13 and that the maximum possible Si in the RIS region consumes the whole Si in matrix then we find the RIS reaches its maximum of 7.5 wt.% at 17 dpa. Within this model no further Si moves to the GB after 17 dpa and, as a result, the γ' volume fraction remains unchanged after 17 dpa.

Using the microalloy methodology described in Sec. 3.2.4 along with CD results for LWR conditions set 3 (see Sec. 3.1.3, Figure 5 and Figure 6) we can predict the evolution of γ' under LWR conditions (Figure 14). For RIS data we use the Si segregation from Figure 13 and for other elements we use the values in Table 5 with linear interpolation up to 17 dpa. We need again to emphasize that the Figure 14 is based on this assumption the segregation to dislocations is similar to experimental segregation data that we have for grain boundaries. While this is clearly an approximate approach, these results suggest that for LWR conditions we might see significantly higher volume fractions of γ' in austenitic steels, approaching 3 %, than seen under accelerated fast reactor tests, which were closer to 0.6% (see Figure 10). The difference is primarily due to the lower temperature condition in light water reactors. In particular, at lower temperatures the Frank loops will form more quickly and lead to larger dislocation density, which increases the volume fraction of RIS regions and nucleation sites for forming γ'.

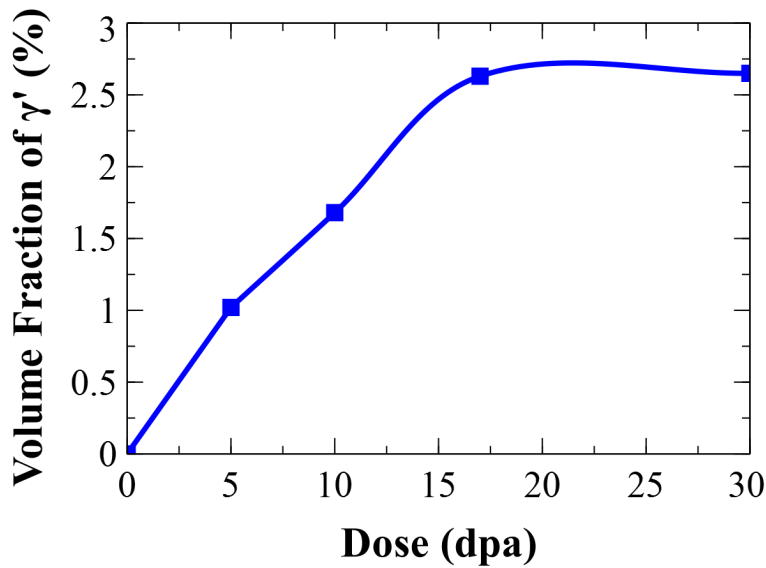

**Figure 14.** The evolution of volume fraction of γ' under LWR conditions (275 °C and $7 \times 10^{-8}$ dpa/s, conditions set 3). The plot is produced based on this assumption that the element segregation to dislocations is similar as that to grain boundaries.

The predicted volume fraction for γ' in Figure 14 is quite high and can potentially cause a considerable hardening in base material. A significant source of uncertainty in our γ' prediction is our simplified RIS model. The volume fraction of γ' in Figure 14 is directly related to Frank loops size and number density, the width of RIS region, and the element concentration in the RIS regions. The CD model is fairly successful in predicting the loop behavior (Figure 5) compared with experimental data. The width of RIS region (4 nm) is also based on experimental data [56].

However, the RIS data we used for predicted γ' comes from RIS experimental observations on grain boundaries under LWR conditions, not from measurements on dislocations. As we are not aware of any experimental data on RIS around the dislocations under LWR conditions, we are forced to assume the RIS behavior at dislocations is similar to grain boundaries. One justification for this assumption is the experimental observations of Jiao and Was [55]. They used the Atom Probe Tomography (APT) technique to study the RIS behavior in 304 SS under 2 MeV proton irradiation to a dose of 5 at 360 °C. They reported that the qualitative RIS behavior (segregation and depletion) of elements at grain boundaries and dislocations were almost identical. More quantitatively, they reported somewhat greater segregation for Fe, Cr, Ni, Mn, and Cu at grain boundaries compared to dislocations. However, for Si it was reverse, and the Si segregation amount was somewhat higher on dislocations than grain boundaries.

We were able to find four papers reporting studies on CW 316 SS under low-flux neutron conditions which could be used for comparison to the prediction in Figure 14 and we discuss each briefly here.

The experimental works on γ' precipitation in CW 316 SS under LWR conditions of which we are aware are by Etienne et al. [61], Edwards et al. [10], Fukuya et al. [62], and Fujii et al. [64]. Etienne et al. [61] performed a chemical analyses (used ATP) on a CW 316 SS bolt irradiated up to 12 dpa at 360 °C and Edwards et al. [10] performed a microstructural characterization (used TEM) on a CW 316 SS bolt irradiated up to 12.2 dpa at 343 °C. Neither Etienne nor Edwards reported the dose rate in their experiments, but we will assume they are similar as both bolts were extracted from a PWR. Fukuya et al. [62] studied the evolution of microstructure and microchemistry by TEM on CW 316 SS up to 73 dpa ($1.5 \times 10^{-7}$ dpa/s and 292-323 °C). Fujii et al. [64] used the APT technique to analyze CW 316 SS specimens irradiated up to 74 dpa in PWR ($1.5 \times 10^{-7}$ dpa/s and 305 °C). Despite the similarity between materials and irradiation conditions, the results were quite different. Etienne et al. [61] reported a volume fraction of 8.6% for Ni-Si clusters, Edwards [10] reported 0.08% volume fraction for γ', Fukuya [62] reported a volume fraction of 0.02% for γ', and Fujii [64] reported 3% volume fraction for Ni-Si clusters. We believe that the difference may come from different temperatures, different irradiation, and/or different experimental techniques, as the APT has a higher capability to detect small clusters than TEM. Etienne [61] and Fujii [64] reported an approximately ten times higher number density for Ni-Si clusters than Edwards [10] and Fukuya [62]. The Ni-Si clusters number density in Etienne and Fujii's work is similar to the Frank loop number density, which is in agreement with other experimental observation of Ni-Si segregation to Frank loops [55, 56].

Edwards et al. in another work [66] compared the microstructural evolution in stainless steels under LWR and fast reactors conditions (used TEM). They reported no precipitation in two CW 316 SS irradiated at BOR-60 (20 dpa, $9\times10^{-7}$ dpa/s, 320 °C) and in a PWR (33 dpa, $0.5\times10^{-7}$ dpa/s, 290 °C). They also reported very small amount of precipitation (number density $< 10^{21}$ m$^{-3}$ and size ~ 10 nm) in CW 316 SS irradiated in a PWR up to 70 dpa at 315 °C with flux $1\times10^{-7}$ dpa/s. They finally concluded that there are some variations in microstructure over the range of dose rates and total accumulated dose, but these changes are relatively minor.

Finally, Allen et al. [45] studied low flux neutron irradiated (~$1 \times 10^{-7}$ dpa/s) CW 316 SS at 375 °C and did not observe any γ' up to 30 dpa.

This data suggests that a high volume fraction of Ni-Si phase precipitates is at least possible, but also provides a few cases where it has not been observed. Considering the limited available experimental data and the level of variation in the reported data, we believe that the accuracy of our model for predicting γ' precipitation under LWR conditions still needs further study. However, our model does suggest that large γ' precipitate concentrations are possible and that this is an important area for further study.

## 4. Summary of model assumptions, approximations, and related sources of error

As there are many assumptions and approximations in the overall modeling approach, and the assumptions depend somewhat on what is being simulated, we include here a summary of the major assumptions for convenience. More detailed sensitivity testing on each of these assumptions would be a valuable area for future work.

1. Cluster dynamics (CD) model approximations
    1.1. The CD model has several input data. Generally, a complete set of input data for a complex alloy like 316 SS is not available. We use available parameters in literature and fit the binding energy between dimer interstitials to reproduce the BOR-60 data. For other irradiation conditions the damage efficiency and in-cascade clustering of defect were fitted to be able to reproduce data over a range of flux and temperatures and compositions.
    1.2. In the CD model we assume that only monomers are mobile. Molecular simulations of some materials, e.g. Ni, have shown that there is a possibility that dimers and trimers be also mobile [67].
    1.3. In the CD model we assume that the interstitials and vacancies clusters are two dimensional (Frank loop).

1.4. In the CD model we assume that the sink effects of precipitates for defects are negligible. This assumption is reasonable based on the low precipitate concentration reported in experimental observations [9, 10, 13].

1.5. In the CD model we assume that the elements diffusion is vacancy mediated and the effect of interstitials on element mobility is negligible.

2. Thermodynamic model (OCTANT) approximations

2.1. OCTANT (thermodynamic database) has some uncertainties for low temperature predictions as no experimental aging data for 316 SS at temperatures lower than 400 °C exist. Similar uncertainties also exist for the mobility database mc_fe_v2.006 [46].

2.2. Minor elements are not currently available in OCTANT. Some of these elements, e.g. Ti, could have direct effect on alloys precipitation behavior.

2.3. We assume that the irradiation does not alter the Gibbs free energy of phases.

2.4. The OCTANT database predicted both ferrite and G-phase, but we do not see either in the experimental observations of most irradiated CW 316 SS. This is most likely due to issues in the kinetics, interfacial energies, or the thermodynamics.

3. Precipitation model (MatCalc) approximations

3.1. We assume all interfacial energies are given by a simple broken bond model, which may impact significantly what dpa subtle transitions occur (e.g., $M_{23}C_6$ to $M_6C$). Large errors could also potentially change the qualitative rates for formation of precipitates from the very rapid (<0.1 dpa for a given state of dislocations and RIS) values found in this work.

3.2. In precipitation modeling we assume that the RIS matrix composition are constant during a given precipitation evolution to its asymptotic volume fraction. These constant values are reasonable assumption as the kinetics of precipitation at both fast reactor and LWR flux, $9.4 \times 10^{-7}$ dpa/s and $7 \times 10^{-8}$ dpa/s, respectively, occurs very quickly (within 0.01 dpa) compared the RIS (several dpa).

3.3. We follow the approximate model for stabilization of precipitates on dislocations built into the MatCalc code that may have some significant errors [68].

3.4. In addition to second phase precipitation in the austenite matrix, it was shown experimentally that austenite might transform to ferrite after high dose irradiation. This phase transformation could lead to alloy embrittlement [69]. It may also effect G-phase precipitation as G-phase nucleation at the boundary of austenite and ferrite was observed in duplex steels [70].

3.5. We assume that the carbon mobility is unchanged under irradiation. However, the carbon-vacancy binding may alter the carbon mobility [71].
4. Radiation-induced segregation (RIS) model approximations
   4.1. In quantitative modeling of γ' we assume that the width of RIS microalloy region is stationary under irradiation (and it is equal to 4 nm [56]).
   4.2. The RIS compositions in Table 5 come from distinct experimental data with different irradiation condition [16]. These data also have high error bar especially for Si, which is the key element in radiation induced precipitation.
   4.3. We use the measured grain boundary RIS and assume the same RIS occurs at dislocations.
   4.4. For high doses where RIS data is not available, we use linear extrapolation of available data. We assume that this extrapolated RIS is bounded by the value at which all Si is in the RIS microalloy region.
5. In γ' precipitation we only consider the RIS on dislocations. This is a reasonable assumption because the dislocations are preferable nucleation sites for γ' [7] and also because the concentration of other sinks, e.g. grain boundaries, is much less than dislocations.

## 5. Conclusion

A thermo-kinetic model was developed to study the second phase precipitation in 300 series austenitic stainless steels, with a focus on CW 316 stainless steels. We compare the model calculated results to data for a fast reactor flux of $9.4 \times 10^{-7}$ dpa/s and temperatures of 390 °C, and predict behavior for LWR conditions of $7 \times 10^{-8}$ dpa/s and 275 °C. The composition used for all models is given in Table 3 and Table 5 (except for Figure 12, which uses the composition in Table 6). The approach integrated a Cluster Dynamics model for radiation enhanced diffusion and dislocation nucleation site parameters, the OCTANT database for thermodynamics, the mc_fe_v2.006 mobility database for kinetic parameters, and MatCalc for solving the precipitation model. The main results of this work are as follows.

1. Fast reactor test conditions ($9.4 \times 10^{-7}$ dpa/s, 390 °C)
   1.1. The model was successful in semi-quantitative prediction of volume fraction of γ' phase and showed discrepancy with carbides that is likely due to the absence of radiation induced dissolution physics in the model and/or carbon depletion at sinks.
   1.2. The results supported the fact that γ' is the dominant phase in microalloy regions enriched in solutes due to RIS at dislocation loops.

1.3. We showed that in the formation of γ' it is the time scale of radiation induced segregation (i.e., the formation of the microalloy region) that controls the γ' formation, not the kinetics of γ' precipitation from the matrix, which takes place over less than ~0.01 dpa once the RIS is established. The steady state RIS takes several dpa to form on an existing dislocation.
1.4. For CW 316 SS with 0.055 wt.% carbon, carbides are predicted to form ~1.2% volume fraction $M_{23}C_6$ within 0.1 dpa, which gets replaced gradually by $M_6C$ while increasing to a final volume fraction of 1.7% at about 23 dpa. These results are fully consistent with high-temperature experimental ageing data. However, the saturation volume fraction in modeling is about one order of magnitude higher than most irradiation experimental data. We propose the radiation enhanced dissolution and/or carbon depletion at sinks (which may lead to reduced carbide nucleation and growth) as the sources of this discrepancy.
1.5. Carbide behavior reaches steady state at about 23 dpa, which suggests that higher doses are unlikely to alter their volume fraction. However, as they are governed by a balance between driving forces for precipitation and dissolution, small changes in flux, temperature, or composition could alter their volume fraction significantly.
2. LWR conditions ($7 \times 10^{-8}$ dpa/s, 275 °C)
    2.1. The model predicted that $M_{23}C_6$ would saturate at the very beginning of irradiation. However, the transformation of $M_{23}C_6$ to $M_6C$ was slower in LWR compared with fast reactors and $M_{23}C_6$ remained the dominant carbide phase up to 100 dpa.
    2.2. In LWR the Frank loops are smaller and more numerous (because of low temperature condition i.e. 275 °C) compared with fast reactors. If we assume similar RIS for dislocation as measure for grain boundaries (Figure 13), we can predict that essentially all Si in 316 SS matrix will segregate to dislocations within 17 dpa.
    2.3. The integrated model along with experimental RIS predicted the maximum volume fraction of γ' to be ~3% under LWR conditions.
    2.4. If radiation enhanced dissolution of carbides and/or segregation of carbon play a critical role in carbide volume fraction in under fast reactor conditions, then lowering the temperature and dose rate to LWR conditions may enhance the carbide volume fraction, potentially increasing values to be up to the thermodynamic prediction of ~1.2%, or an approximately ten-fold increase vs. values under fast reactor conditions.

This work has also led to a modeling that can be readily adapted to new temperature, flux, and fluence conditions in austenitic stainless steels. Through our focus on CW 316 SS we have developed a new understanding of the time scales and mechanisms governing the formation and evolution under irradiation of the key precipitate phases: carbides, γ', and G-phase.

**Acknowledgement**

We gratefully acknowledge funding from the Department of Energy Light Water Reactor Sustainability (LWRS) Program.

**References**

[1] S.J. Zinkle, J.T. Busby. Structural materials for fission & fusion energy, Materials Today 12 (2009) 12-19.
[2] T. Allen, J. Busby. Radiation damage concerns for extended light water reactor service, JOM 61 (2009) 29-34.
[3] E.A. Kenik, J.T. Busby. Radiation-induced degradation of stainless steel light water reactor internals, Materials Science and Engineering: R: Reports 73 (2012) 67-83.
[4] P. Maziasz. Overview of microstructural evolution in neutron-irradiated austenitic stainless steels, Journal of nuclear materials 205 (1993) 118-145.
[5] P. Maziasz, C. McHargue. Microstructural evolution in annealed austenitic steels during neutron irradiation, International materials reviews 32 (1987) 190-219.
[6] E. Lee, L. Mansur. Fe–15Ni–13Cr austenitic stainless steels for fission and fusion reactor applications. III. Phase stability during heavy ion irradiation, Journal of nuclear materials 278 (2000) 20-29.
[7] E. Lee, P. Maziasz, A. Rowcliffe. Structure and composition of phases occurring in austenitic stainless steels in thermal and irradiation environments. Oak Ridge National Lab., TN (USA), 1980.
[8] W. Yang, H. Brager, F. Garner. Radiation-induced phase development in AISI 316. Hanford Engineering Development Lab., Richland, WA (USA), 1980.
[9] D. Edwards, A. Schemer-Kohrn, S. Bruemmer. Characterization of neutron-irradiated 300-series stainless steels, EPRI, Palo Alto, CA 1009896 (2006).
[10] D.J. Edwards, E.P. Simonen, F.A. Garner, L.R. Greenwood, B.M. Oliver, S.M. Bruemmer. Influence of irradiation temperature and dose gradients on the microstructural evolution in neutron-irradiated 316SS, Journal of nuclear materials 317 (2003) 32-45.
[11] A.R. Laborne, P. Gavoille, J. Malaplate, C. Pokor, B. Tanguy. Correlation of radiation-induced changes in microstructure/microchemistry, density and thermo-electric power of type 304L and 316 stainless steels irradiated in the Phénix reactor, Journal of Nuclear Materials 460 (2015) 72-81.
[12] A. Renault-Laborne, J. Malaplate, C. Pokor, B. Tanguy. Characterization of Precipitates in 316 Stainless Steel Neutron-Irradiated at 390 C by the Combination of CDF-TEM, EF-TEM, and HR-TEM. In: Kirk M, Lucon E, (Eds.). Effects of Radiation on Nuclear Materials, ASTM STP 1572. West Conshohocken, PA: ASTM International, 2014. p.1–24.
[13] L. Tan, K.G. Field, J.T. Busby. Analysis of Phase Transformation Studies in Solute Addition Alloys, ORNLTM-2014303 (2014).
[14] Y. Yang, J. Busby. Thermodynamic modeling and kinetics simulation of precipitate phases in AISI 316 stainless steels, Journal of Nuclear Materials 448 (2014) 282-293.
[15] E. Kozeschnik, B. Buchmayr. MATCALC- a simulation tool for multicomponent thermodynamics, diffusion and phase transformations. Fifth International Seminar on the Numerical Analysis of Weldability, 1999. p.349-361.
[16] J.-H. Shim, E. Povoden-Karadeniz, E. Kozeschnik, B.D. Wirth. Modeling precipitation thermodynamics and kinetics in type 316 austenitic stainless steels with varying composition as an initial step toward predicting phase stability during irradiation, Journal of Nuclear Materials 462 (2015) 250-257.

[17]     R. Kampmann, R. Wagner. Decomposition of alloys: the early stages. Proc. 2nd Acta-Scripta Metall. Conf., Pergamon, Oxford, 1984. p.91-103.
[18]     J. Svoboda, F. Fischer, P. Fratzl, E. Kozeschnik. Modelling of kinetics in multi-component multi-phase systems with spherical precipitates: I: Theory, Materials Science and Engineering: A 385 (2004) 166-174.
[19]     L. Onsager. Reciprocal relations in irreversible processes. I, Physical Review 37 (1931) 405.
[20]     L. Onsager. Reciprocal relations in irreversible processes. II, Physical Review 38 (1931) 2265.
[21]     E. Kozeschnik. Modeling Solid-State Precipitation, Momentum Press, 2012.
[22]     S.D. Cohen, A.C. Hindmarsh. CVODE, a stiff/nonstiff ODE solver in C, Computers in physics 10 (1996) 138-143.
[23]     F. Christien, A. Barbu. Effect of self-interstitial diffusion anisotropy in electron-irradiated zirconium: A cluster dynamics modeling, Journal of nuclear materials 346 (2005) 272-281.
[24]     A.H. Duparc, C. Moingeon, N. Smetniansky-de-Grande, A. Barbu. Microstructure modelling of ferritic alloys under high flux 1 MeV electron irradiations, Journal of nuclear materials 302 (2002) 143-155.
[25]     A. Gokhman, F. Bergner. Cluster dynamics simulation of point defect clusters in neutron irradiated pure iron, Radiation Effects & Defects in Solids: Incorporating Plasma Science & Plasma Technology 165 (2010) 216-226.
[26]     C. Pokor, Y. Brechet, P. Dubuisson, J.P. Massoud, A. Barbu. Irradiation damage in 304 and 316 stainless steels: experimental investigation and modeling. Part I: Evolution of the microstructure, Journal of Nuclear Materials 326 (2004) 19-29.
[27]     J. Ribis, F. Onimus, J.-L. Béchade, S. Doriot, A. Barbu, C. Cappelaere, C. Lemaignan. Experimental study and numerical modelling of the irradiation damage recovery in zirconium alloys, Journal of Nuclear Materials 403 (2010) 135-146.
[28]     D. Brimbal, L. Fournier, A. Barbu. Cluster dynamics modeling of the effect of high dose irradiation and helium on the microstructure of austenitic stainless steels, Journal of Nuclear Materials 468 (2016) 124-139.
[29]     R. Stoller. Modeling dislocation evolution in irradiated alloys, Metallurgical Transactions A 21 (1990) 1829-1837.
[30]     W. Phythian, R. Stoller, A. Foreman, A. Calder, D. Bacon. A comparison of displacement cascades in copper and iron by molecular dynamics and its application to microstructural evolution, Journal of nuclear materials 223 (1995) 245-261.
[31]     Y.N. Osetsky, A. Serra, M. Victoria, S. Golubov, V. Priego. Vacancy loops and stacking-fault tetrahedra in copper: I. Structure and properties studied by pair and many-body potentials, Philosophical Magazine A 79 (1999) 2259-2283.
[32]     N. Soneda, T.D. De La Rubia. Defect production, annealing kinetics and damage evolution in α-Fe: an atomic-scale computer simulation, Philosophical Magazine A 78 (1998) 995-1019.
[33]     R.E. Stoller, G.R. Odette. A composite model of microstructural evolution in austenitic stainless steel under fast neutron irradiation. In: F. A. Garner NHP, A. S. Kumar, (Ed.). Radiation-Induced Changes in Microstructure, 13th International Symposium, ASTM STP 955. Seattle, WA, 1987. p.371-392.
[34]     D. Gelles. A Frank loop unfaulting mechanism in fcc metals during neutron irradiation. Dislocation modelling of physical systems. 1981.

[35] N. Ghoniem, D. Cho. The simultaneous clustering of point defects during irradiation, physica status solidi (a) 54 (1979) 171-178.
[36] A. Barashev, S. Golubov, R. Stoller. A Model of Radiation-induced Microstructural Evolution, ORNLLTR-2014487 (2014).
[37] J. Gan, G. Was, R. Stoller. Modeling of microstructure evolution in austenitic stainless steels irradiated under light water reactor condition, Journal of nuclear materials 299 (2001) 53-67.
[38] P. Ehrhart, P. Jung, H. Schultz, H. Ullmaier. Atomic Defects in Metals, Springer, 1991.
[39] G. Odette, T. Yamamoto, D. Klingensmith. On the effect of dose rate on irradiation hardening of RPV steels, Philosophical Magazine 85 (2005) 779-797.
[40] A. Renault-Laborne, J. Garnier, J. Malaplate, P. Gavoille, F. Sefta, B. Tanguy. Evolution of microstructure after irradiation creep in several austenitic steels irradiated up to 120 dpa at 320° C, Journal of Nuclear Materials 475 (2016) 209-226.
[41] Y. Yang, L. Tan, J.T. Busby. Thermal Stability of Intermetallic Phases in Fe-rich Fe-Cr-Ni-Mo Alloys, Metallurgical and Materials Transactions A 46 (2015) 3900-3908.
[42] L. Tan, Y. Yang. In situ phase transformation of Laves phase from Chi-phase in Mo-containing Fe–Cr–Ni alloys, Materials Letters 158 (2015) 233-236.
[43] A.-É. Renault, C. Pokor, J. Garnier, J. Malaplate. Microstructure and grain boundary chemistry evolution in austenitic stainless steels irradiated in the BOR-60 reactor up to 120 dpa, . 14th Int Conf. on Environmental Degradation of Materials in Nuclear Power Systems, Virginia Beach, VA, 2009. p.1324-1334.
[44] N. Hashimoto, E. Wakai, J. Robertson, T. Sawai, A. Hishinuma. Microstructure of austenitic stainless steels irradiated at 400 C in the ORR and the HFIR spectral tailoring experiment, Journal of nuclear materials 280 (2000) 186-195.
[45] T.R. Allen, H. Tsai, J.I. Cole, J. Ohta, K. Dohi, H. Kusanagi. Properties of 20% Cold-Worked 316 Stainless Steel Irradiated at Low Dose Rate, Effects of Radiation on Materials, 21th International Symposium, ASTM STP 1447 (2004) p. 1-12.
[46] E. Povoden-Karadeniz. MatCalc Mobility Database "mc_fe_V2.006". Vienna University of Technology.
[47] B. Sonderegger, E. Kozeschnik. Generalized nearest-neighbor broken-bond analysis of randomly oriented coherent interfaces in multicomponent fcc and bcc structures, Metallurgical and Materials Transactions A 40 (2009) 499-510.
[48] P. Maziasz. The formation diamond-cubic eta (η) phase in type 316 stainless steel exposed to thermal aging or irradiation environments, Scripta Metallurgica 13 (1979) 621-626.
[49] M.N. Gussev, J.T. Busby, L. Tan, F.A. Garner. Magnetic phase formation in irradiated austenitic alloys, Journal of Nuclear Materials 448 (2014) 294-300.
[50] B. Margolin, I. Kursevich, A. Sorokin, V. Neustroev. The Relationship of Radiation Embrittlement and Swelling for Austenitic Steels for WWER Internals. ASME 2009 Pressure Vessels and Piping Conference: American Society of Mechanical Engineers, 2009. p.939-948.
[51] B. Weiss, R. Stickler. Phase instabilities during high temperature exposure of 316 austenitic stainless steel, Metallurgical Transactions 3 (1972) 851-866.
[52] J. Lai. A study of precipitation in AISI type 316 stainless steel, Materials Science and Engineering 58 (1983) 195-209.
[53] National Institute for Materials Science (NIMS). Diffusion Database. vol. 2015.
[54] H.J. Goldschmid. Interstitial alloys, Springer, 2013.

[55] Z. Jiao, G. Was. Novel features of radiation-induced segregation and radiation-induced precipitation in austenitic stainless steels, Acta Materialia 59 (2011) 1220-1238.
[56] M. Hatakeyama, S. Tamura, I. Yamagata. Direct observation of solute–dislocation interaction on screw dislocation in a neutron irradiated modified 316 stainless steel, Materials Letters 122 (2014) 301-305.
[57] K.C. Russell. Phase stability under irradiation, Progress in Materials Science 28 (1984) 229-434.
[58] R. Nelson, J. Hudson, D. Mazey. The stability of precipitates in an irradiation environment, Journal of Nuclear Materials 44 (1972) 318-330.
[59] E. Lee, L. Mansur. Relationships between phase stability and void swelling in Fe-Cr-Ni alloys during irradiation, Metallurgical Transactions A 23 (1992) 1977-1986.
[60] C. Cawthorne, C. Brown. The occurrence of an ordered FCC phase in neutron irradiated M316 stainless steel, Journal of Nuclear Materials 66 (1977) 201-202.
[61] A. Etienne, B. Radiguet, P. Pareige, J.-P. Massoud, C. Pokor. Tomographic atom probe characterization of the microstructure of a cold worked 316 austenitic stainless steel after neutron irradiation, Journal of Nuclear Materials 382 (2008) 64-69.
[62] K. Fukuya, K. Fujii, H. Nishioka, Y. Kitsunai. Evolution of microstructure and microchemistry in cold-worked 316 stainless steels under PWR irradiation, Journal of nuclear science and technology 43 (2006) 159-173.
[63] G. Bond, B. Sencer, F. Garner, M. Hamilton, T. Allen, D. Porter. Void Swelling of Annealed 304 Stainless Steel at∼ 370-385° C and PWR-Relevant Displacement Rates. In: Ford FP, Bruemmer SM, Was GS, (Eds.). 9th International Symposium on Environmental Degradation of Materials in Nuclear Power Systems-Water Reactors, TMS (The Minerals, Metals & Materials Society), 1999. p.1045-1050.
[64] K. Fujii, K. Fukuya. Irradiation-induced microchemical changes in highly irradiated 316 stainless steel, Journal of Nuclear Materials 469 (2016) 82-88.
[65] S.M. Bruemmer, E.P. Simonen, P.M. Scott, P.L. Andresen, G.S. Was, J.L. Nelson. Radiation-induced material changes and susceptibility to intergranular failure of light-water-reactor core internals, Journal of Nuclear Materials 274 (1999) 299-314.
[66] D. Edwards, E. Simonen, S. Bruemmer, P. Efsing. Microstrutural evolution in neutron irradiated stainless steels: Comparison of LWR and fast reactors irradiation. In: Allen TR, King PJ, Nelson L, (Eds.). 12th International Conference on Environmental Degradation of Materials in Nuclear Power System – Water Reactors, TMS (The Minerals, Metals & Materials Society), 2005. p.419-428.
[67] N. Lam, L. Dagens. Calculations of the properties of single and multiple defects in nickel, Journal of Physics F: Metal Physics 16 (1986) 1373.
[68] R. Radis, E. Kozeschnik. Numerical simulation of NbC precipitation in microalloyed steel, Modelling and Simulation in Materials Science and Engineering 20 (2012) 055010.
[69] B. Margolin, E. Yurchenko, A. Morozov, D. Chistyakov. Prediction of the effects of thermal ageing on the embrittlement of reactor pressure vessel steels, Journal of Nuclear Materials 447 (2014) 107-114.
[70] T. Hamaoka, A. Nomoto, K. Nishida, K. Dohi, N. Soneda. Accurate determination of the number density of G-phase precipitates in thermally aged duplex stainless steel, Philosophical Magazine 92 (2012) 2716-2732.
[71] J. Slane, C. Wolverton, R. Gibala. Experimental and theoretical evidence for carbon-vacancy binding in austenite, Metallurgical and Materials Transactions A 35 (2004) 2239-2245.